\documentclass[11 pt]{article}
\usepackage{threeparttable}
\usepackage{amsmath}
\usepackage{amssymb}
\usepackage[a4paper, total={6.5in, 9in}]{geometry}
\usepackage{amsthm,url,tikz}
\usepackage[utf8]{inputenc}
\usepackage[english]{babel}
\usepackage{graphicx}
\usepackage{hyperref}
\usepackage{adjustbox}
\numberwithin{equation}{section}
\usepackage{makecell}
\usepackage{bm}
\usepackage{xcolor}
\newtheorem{theorem}{Theorem}[section]
\newtheorem{proposition}[theorem]{Proposition}
\newtheorem{corollary}[theorem]{Corollary}
\newtheorem{lemma}[theorem]{Lemma}
\newtheorem{definition}[theorem]{Definition}

\newtheorem{example}[theorem]{Example}
\newtheorem{remark}[theorem]{Remark}
\usepackage{enumerate}
\usepackage[nottoc,notlot,notlof]{tocbibind} 
\usepackage{mathtools}
\usepackage{amsthm}
\usepackage{booktabs}
\usepackage[backend=biber, style=ieee, sorting=nyt]{biblatex}
\usepackage{multirow}
\begin{document}
\title{Function-Correcting Codes for Insertion-Deletion Channels}
%
%
%

\author{ 
        Anamika Singh and
        Abhay Kumar Singh
\thanks{ A. Singh and A. K. Singh are with the Department of Mathematics and Computing, Indian Institute of Technology (ISM), Dhanbad, India. email: anamikabhu2103@gmail.com, abhay@iitism.ac.in}}

\maketitle

\begin{abstract}

In coding theory, handling errors introduced by symbol insertion or deletion in a transmitted message is a long-standing challenge. Optimizing redundancy for insertion and deletion channels remains a key open problem, with significant importance for applications in DNA data storage and document exchange. Recently, a new coding framework, function-correcting codes, has been proposed to address the challenge of optimizing redundancy while preserving specific message functions. This framework has attracted attention for its potential applications in machine learning systems and in long-term archival data storage. To address the problem of redundancy optimization in insertion-deletion channels, we propose a new coding framework, function-correcting codes for insertion-deletion channels.

In this paper, we introduce the notions of function-correcting insertion codes, function-correcting deletion codes, and function-correcting insdel (insertion–deletion) codes (FCIDCs), and we demonstrate that these three formulations are equivalent. We then introduce insdel-distance matrices and irregular insdel-distance codes, and further derive lower and upper bounds on the optimal redundancy achievable by function-correcting codes for insdel channels. Furthermore, we establish Gilbert-Varshamov and Plotkin-like bounds on the length of irregular insdel-distance codes. By utilizing the relation between optimal redundancy and the length of irregular insdel-distance codes, we provide another simplified lower bound on optimal redundancy. We subsequently find bounds on optimal redundancy of FCIDCs for various classes of functions, including locally bounded functions, VT syndrome functions, the number-of-runs function, and the maximum-run-length function.
\end{abstract}

\textbf{Keywords}:
 Function-correcting codes, error-correcting codes, insertion-deletion channels, optimal redundancy.

%

\section{Introduction}
%
%
%
%

{T}{raditional} error-correcting codes (ECCs) are designed to enable the decoder to recover the transmitted message exactly. However, in many real-world situations, the decoder simply needs to compute a certain function of the message rather than reconstructing the complete message. The authors of \cite{10132545} were inspired by this realization to develop a new class of codes called function-correcting codes (FCCs), which encode the message so that the decoder can reliably compute the desired function value, even in the presence of errors, with substantially less redundancy than required by classical ECCs for full message recovery.

FCCs have been commonly studied under systematic encoding, which is important in applications such as distributed computing and long-term archival storage, where preserving the original data is essential. Since redundancy is the primary and most crucial factor for FCCs, it becomes important to precisely determine optimal redundancy—that is, the minimum number of additional bits required to guarantee correct computation of the intended function. To determine optimal redundancy for both generic and specific functions, the authors of \cite{10132545} introduced the concept of irregular-distance codes, which are closely connected to the redundancy patterns of FCCs. They established equivalence between the optimal redundancy of FCCs and the shortest achievable length of irregular-distance codes. Using this equivalence, the authors derived redundancy bounds for arbitrary functions and subsequently used these results to obtain tight or near-tight bounds for several important function classes, including locally binary bounded functions, Hamming weight functions, Hamming weight distribution functions, and real-valued functions.

 \subsection{Motivation}
    The nature of the underlying channel noise fundamentally determines how function values are distorted during transmission. A function that is robust to substitution errors may behave unpredictably under insertions and deletions, since these errors disrupt the alignment between transmitted and received sequences rather than merely altering symbol values. By designing FCCs tailored to specific channel models, one can achieve reliable function computation in scenarios where recovering the entire message is either unnecessary or prohibitively costly in redundancy, while also gaining deeper insight into how different error processes affect function evaluation. Motivated by this observation, we focus on insertion–deletion (insdel) channels, where errors occur when symbols are inserted into or deleted from a transmitted message. Such errors create a form of noise that traditional error-correcting codes, designed primarily for substitution errors, are not well-suited to handle, and they frequently occur in biological systems, modern storage devices, and asynchronous communication systems.

    The study of insertion–deletion errors began in the 1960s \cite{varshamov1965codes,levenshtein1965двоичные, levenshtein1967asymptotically, tenengolts1984nonbinary}, when Levenshtein introduced the edit distance, defining the minimum number of insertions, deletions, or substitutions required to transform one string into another. This concept established the mathematical basis for analyzing synchronization errors. Later, practical coding schemes for handling such errors began to emerge, most notably the Varshamov–Tenengolts (VT) codes \cite{varshamov1965codes}, which enable reliable single-deletion correction, alongside the use of synchronizing sequences to maintain alignment between transmitted and received data.

    The practical importance of function correction under the insdel metric is most evident in DNA-based data storage. In such systems, information is encoded into synthetic DNA strands and recovered through sequencing, a process in which insertions and deletions are among the dominant error types. Consequently, the underlying communication model is naturally described by the insertion--deletion metric rather than the substitution-based Hamming metric. In many storage and retrieval pipelines, the decoder does not need to reconstruct an entire strand but only a function of it, such as a synchronization marker, an addressing label used to identify a strand, or a constraint-related property employed during decoding and error detection. Protecting only the function value, rather than the entire message, can substantially reduce the required redundancy and hence the synthesis cost, which is widely regarded as one of the principal bottlenecks in DNA-based storage systems. The specific functions considered in this work are closely related to this setting. The number-of-runs and maximum-run-length functions capture homopolymer characteristics that directly affect DNA synthesis and sequencing reliability, while the VT-syndrome function forms the basis of classical single insertion--deletion error correction. Similar requirements also arise in file synchronization and document exchange, where modifications naturally appear as insertions and deletions and one may wish to reliably recover a checksum, fingerprint, or version identifier rather than reconstruct the entire file.

    Consequently, the development of function-correcting codes under the insdel metric represents both a substantial theoretical milestone and a vital practical capability for next-generation communication and computation systems. These codes extend conventional error-control techniques beyond the Hamming metric by providing resilience against synchronization errors, enabling more reliable information processing, transmission, and storage in emerging high-density, high-throughput environments such as DNA archival storage.
    
 \subsection{Related Works}
    Function-correcting codes (FCCs) were introduced by Lenz et al. in \cite{10132545} as a coding framework for recovering the value of a target function from the data rather than the data itself, resulting in substantially lower redundancy than traditional error-correcting codes. Their work formalized FCCs, established an equivalence with irregular-distance codes, and provided general upper and lower bounds demonstrating how the structure of the protected function influences the required redundancy. Subsequent developments, such as \cite{Xia2023}, revisited the model and extended the framework from the Hamming metric to the symbol-pair metric, deriving several bounds and structural insights. In \cite{Premlal2024}, the authors established a lower bound on the redundancy of FCCs for linear functions. Focusing on the Hamming weight and its distribution, Ge et al. \cite{Ge2025} established lower bounds on redundancy and introduced code designs that exactly meet those bounds.

    A growing line of research has further generalized FCCs to non-standard error models. In particular, \cite{Singh2025} extended function-correcting codes to the $b$-symbol read channel, addressing clustered read errors common in modern storage systems. That work introduced irregular $b$-symbol distance codes and derived bounds characterizing the redundancy needed to recover a function value under this metric. The results, supported by a graph-theoretic formulation and illustrative examples, show that FCCs can achieve substantially lower redundancy than traditional $b$-symbol error-correcting codes when only the function output must be protected.

    Further extensions were provided in \cite{11250740}, which studied function-correcting codes for locally bounded $b$-symbol functions over $b$-symbol read channels. Recently, function correction has also been explored in the Lee-metric setting \cite{rajan2025plotkin, Verma2025FCCL}. More recent work, such as \cite{rajput2025function}, investigates hybrid protection models in which both the function value and selected parts of the data must remain resilient to errors, highlighting the versatility of FCCs as a targeted and efficient alternative to full-message protection.

    \subsection{Contributions}

    Existing work on function-correcting codes has been developed almost entirely for substitution-type metrics such as the Hamming, symbol-pair, and Lee metrics, in which a received symbol may be wrong, but symbol positions remain aligned. A defining feature of all these metrics is that the underlying distance is additive across a systematic split, so the message and redundancy blocks decouple, and the irregular-distance machinery of \cite{10132545} applies directly. The insertion-deletion metric does not share this property. Cross-block alignment between concatenated words can reduce the insdel distance of a concatenation, so the two blocks no longer decouple. This single phenomenon undermines the mechanism on which the substitution-type metrics theory rests and requires the use of a different set of tools. The following are the contributions of this paper.

    \begin{itemize}

    \item We introduce function-correcting deletion, insertion, and insdel codes and prove that the three formulations coincide.

    \item To deal with the cross-block LCS loss, we introduce two types of insdel-distance matrices. We further show that this loss confines nontrivial systematic function-correcting insdel codes to the regime $r\ge k$. For $r<k$, the message vectors must already form a classical $t$-insdel code, and the framework offers no savings. We also define irregular insdel-distance codes, relate their lengths to the optimal redundancy $r^{f}_{ID}(k,t)$, and establish Plotkin-like and Gilbert--Varshamov bounds for them. Because insdel balls are center-dependent, their size depends on the number-of-runs of the center. These bounds rely on supersequence and run-based arguments rather than the uniform sphere-counting of the Hamming case.

    \item We derive bounds and explicit codes for the insdel-native functions, such as the VT-syndrome function, the number-of-runs function, the maximum-run-length function, and the locally bounded functions. We also show that when $f$ is bijective, the framework reduces to classical $t$-insdel error correction, and its bounds specialize accordingly, providing a consistency verification against known insdel bounds.

    \end{itemize}
    \subsection{Organization}
    The rest of the paper is organized as follows. Section II provides a comprehensive review of the foundational concepts used throughout the paper. We begin by recalling the basic definitions and properties of insertion–deletion (insdel) codes, including the insdel metric and several equivalent formulations that are relevant to our analysis. In Section III, we introduce the framework of function-correcting deletion codes, function-correcting insertion codes, and function-correcting insertion–deletion (insdel) codes, which are designed to recover the value of a target function in the presence of deletion, insertion, or combined insertion–deletion errors. We then present and discuss irregular insdel-distance codes, highlighting their significance in characterizing optimal redundancy and in deriving key performance limits. Building on this framework, we derive lower and upper bounds on the optimal redundancy achievable by function-correcting codes for insdel channels. In addition, we establish Gilbert–Varshamov-type and Plotkin-like bounds on the lengths of irregular insdel-distance codes, which play a central role in analyzing and understanding the redundancy of such codes. In Section IV, we apply the general theoretical results to a specific class of functions, the VT syndrome function, which plays a central role in the insertion–deletion coding. We analyze how the structure of the VT syndrome function interacts with insdel errors and derive both upper and lower bounds on the optimal redundancy required for function correction in this setting. We illustrate the general framework using three concrete classes of functions: the number-of-runs function (Section V), the maximum-run-length function (Section VI), and locally bounded functions (Section VII). For each class, we derive the corresponding redundancy values and their associated upper and lower bounds. Section VIII concludes the paper.   
    \subsubsection{Notations}
    Throughout this paper, we use the following notation:
    \begin{itemize}
    \item $\mathbb{F}_2$: the binary field.
    \item $\mathbb{F}_2^n$: the vector space of all binary vectors of length $n$.
    \item $\mathbb{N}$: the set of positive integers; $\mathbb{N}_0 = \mathbb{N} \cup \{0\}$: the set of non-negative integers.
    \item For vectors $x,y$:
    \begin{itemize}
        \item $d_H(x,y)$: the Hamming distance,
        \item $d_{ID}(x,y)$: the insertion-deletion distance,
        \item $LCS(x,y)$: the length of the longest common subsequence of $x$ and $y$.
        \item $SCS(x,y)$: the length of the shortest common supersequence of $x$ and $y$.
    \end{itemize}
    \item $|x|$: the length of a vector (or string) $x$.
    \item $D_t(x)$: the set of all subsequences obtained by deleting exactly $t$ symbols from $x$.
    \item $I_t(x)$: the set of all supersequences obtained by inserting exactly $t$ symbols into $x$.
    \item $r(x)$: the number-of-runs in a binary vector $x$.
    \item Code parameters:
    \begin{itemize}
        \item $k$: message length,
        \item $r$: redundancy,
        \item $n = k + r$: total codeword length.
    \end{itemize}
    \item $\mathbb{N}_0^{M \times M}$: the set of all $M \times M$ matrices with non-negative integer entries.
    \item $[I]_{ij}$: the $(i,j)$-th entry of a matrix $I$.
    \item For any integer $M$:
    \begin{itemize}
        \item $[a] \triangleq \{1, 2, \dots, a\}$,
        \item $[a]^+ \triangleq \max\{a, 0\}$.
    \end{itemize}
    \item $e$: Euler's number, base of natural logarithm ( $e \, \approx \, 2.71828$).
\end{itemize}
 
\section{Preliminaries}
    In this section, we review the basic concepts and definitions used throughout the paper. We begin with a brief discussion of insertion, deletion, and insdel codes, which are designed to enable reliable communication over channels that produce synchronization errors. We then introduce the fundamental notions associated with irregular-distance codes, a class of codes characterized by non-uniform or structure-dependent distance measures. Throughout the paper, we focus on codes over the alphabet $\mathbb{F}_2 = \{0,1\}$.
 \subsection{Insdel code}
     We consider a channel model in which errors occur as deletions, insertions, or both, and recall the classical codes introduced for this setting in the 1960s (see Section~I) together with the main results we use later.

    Let $x$ be a binary vector of length $k$ that, when transmitted through a deletion channel, may lose some of its bits. 
    If the channel deletes $t$ bits, the resulting binary vector $y$ has length $k - t$. 
    The vector $y$ is referred to as a \textit{subsequence} of $x$. 
    A subsequence of a vector is obtained by selecting a subset of its symbols and aligning them in their original order, 
    without any rearrangement. 
    Formally, the subsequence is defined as follows:

    \begin{definition}[Subsequence]
        A sequence $x = x_1 \ldots x_k$ is called a subsequence of $y = y_1 \ldots y_n$ if there are $k$ indices $i_1 < \ldots < i_k$ such that $x_1 = y_{i_1} \ldots x_k = y_{i_k}.$
    \end{definition}

    Similarly, when $x$ is transmitted through an insertion channel, additional bits may be inserted at arbitrary positions, 
    resulting in a \textit{supersequence} of $x$. 
    In the deletion case, the original vector $x$ is referred to as a supersequence of the received vector $y$.
    The set of all subsequences of $x$ obtained by deletion of $t$ bits is denoted by $D_t(x)$ and the set of supersequences obtained by insertion of $t$ bits is denoted by $I_t(x)$.\\
    We now define the main object of interest, which is \textit{deletion-correcting code} and \textit{insertion-correcting code}.

    \begin{definition}[Deletion Correcting Codes]\cite{levenshtein1965двоичные}
        A $t$-deletion-correcting code $\mathcal{C}$ of length $n$ is a subset of binary vector space $\mathbb{F}_2^n$ that holds the following property for all vectors $x,y \in \mathcal{C}$,
        $$D_t(x) \cap D_t(y) = \emptyset.$$     
    \end{definition}
        
    \begin{definition}[Insertion Correcting Codes]\cite{levenshtein1965двоичные}
        A $t$-insertion correcting code $\mathcal{C}$ of length $n$ is a subset of binary vector space $\mathbb{F}_2^n$ that holds the following property for all vectors $x,y \in \mathcal{C}$,
        $$I_t(x) \cap I_t(y) = \emptyset.$$
    \end{definition}
   Codes that are capable of correcting both insertions and deletions are referred to as \textit{insertion-deletion codes} (or \textit{insdel codes}). These codes can also be characterized through a suitable metric description, defined as follows:
 
    \begin{definition}[Insdel Metric]\cite{levenshtein1965двоичные}
        Let \( x \) and \( y \) be binary sequences of length \( k \). The insdel metric \( d_{ID}(x, y) \) between \( x \) and \( y \) is defined as the minimum number of insertions and deletions required to transform \( x \) into \( y \).
    \end{definition}
       
           
    A sequence $z$ is a common subsequence of $x$ and $y$ if it is a subsequence of both $x$ and $y$. A longest common subsequence is a common subsequence of maximum possible length. Similarly, a sequence $w$ is a common supersequence of $x$ and $y$ if both $x$ and $y$ are subsequences of $w$, and a shortest common supersequence is a common supersequence of minimum possible length.
    Considering only insertion and deletion operations, we can define the insdel distance in terms of the length of the longest common subsequence as follows:
    \begin{equation*}
        d_{ID}(x, y) = |x| + |y| - 2LCS(x,y) = 2k - 2LCS(x,y),
    \end{equation*}
     where $|x|$ denotes the length of a binary word $x$ and $LCS(x,y)$ denotes the length of the longest common subsequence between $x$ and $y$.

     \begin{example}
        Let $x = 101$ and $y = 010$, then the insdel distance between $x$ and $y$ is given as follows:
        \begin{equation*}
        101 \xrightarrow[\text{at position 1}]{deletion} 01 \xrightarrow[\text{at position 3}]{insertion} 010.        
        \end{equation*}
       Therefore, $d_{ID}(x,y) = 2.$ 
    \end{example}
    
    \begin{definition}{(Insdel Code)}\cite{do2021explicit}
        A subset of $\mathbb{F}_2^k$ is defined to be an insdel code $\mathcal{C}$ whose minimum distance is given by:
        $$ d_{ID}(\mathcal{C}) = min_{c_1, c_2 \in \mathcal{C}, c_1 \neq c_2}{d_{ID}(c_1,c_2)}.$$
    \end{definition}
    
     The insdel distance of a code is a key parameter, as it determines the code’s capability to correct insdel errors. A code is said to be a $t$-insdel error-correcting code if its insdel distance is at least $2t+1$.

    \begin{remark}\cite{levenshtein1965двоичные}
    Any code that can correct \(s\) deletions (or equivalently, any code that can correct \(s\) insertions) can also correct \(s\) combined deletions and insertions.
    \end{remark}
    
    The following lemma gives a relation between the Hamming distance and the insdel distance.

    \begin{lemma}\label{lem:rel_insdel_hamming}\cite{liu2022bounds}
    Let $x, y$ be two binary words of length $k$. If $d_H(x,y)$ denotes the Hamming distance between $x$ and $y$ then we have:
    $$ d_{ID}(x,y) \leq 2d_H(x,y).$$
    \end{lemma}

    Insertion and deletion errors can disrupt the alternating structure of a binary string. A useful combinatorial invariant for studying such transformations is the number-of-runs, which we define as follows.

    \begin{definition}[Run and number-of-runs]
    \label{def:run}
    A run in a binary vector $x = (x_1, x_2, \ldots, x_n) \in \mathbb{F}_2^n$ is a maximal contiguous block of identical symbols, that is, a maximal interval of indices $[i, j] \subseteq \{1, \ldots, n\}$ such that $x_i = x_{i+1} = \cdots = x_j$. The number-of-runs in $x$, is the number of such maximal blocks. 
    \end{definition}
    
    For example, the vector $x = 00110100$ partitions into the runs $00$, $11$, $0$, $1$, $00$, yielding $r(x) = 5$.
    The following lemma gives a lower bound on the insdel distance in terms of the difference in number-of-runs. This bound will later help us establish limits on code correction and code construction capabilities.
    \begin{lemma}\label{run_insdel_bound}
        Let $x,y \in \mathbb{F}_2^n$ be two binary vectors of length $n$ and let $r(x)$ and $r(y)$ be the number-of-runs corresponding to vectors $x$ and $y$ respectively. Then,
        $$ d_{ID}(x,y) \ge 2 \Big\lceil \frac{|r(y) - r(x)|}{2} \Big\rceil.$$
        Moreover, when $|r(x) - r(y)|$ is odd then $d_{ID}(x,y)\geq |r(x) - r(y)| + 1.$
    \end{lemma}
    \begin{proof}
    Assume without loss of generality that $r(y) \ge r(x)$. Let $m$ denote the number of insertions (and hence also the number of deletions) in an optimal insertion-deletion transformation from $x$ to $y$, so that the total number of operations is $d_{ID}(x,y)=2m$.

    Let $A$ be the total change in run count caused by the $m$ insertions, and let $B$ be the total change in run count caused by the $m$ deletions. Then
    \[
        r(y) - r(x) = A + B.
    \]

    Since a deletion can never increase the number-of-runs in a binary string, we have $B \le 0$. Moreover, each insertion can increase the number-of-runs by at most $2$, and hence
    \[
        A \le 2m.
    \]

    Combining these inequalities yields
    \[
    r(y) - r(x) = A + B \le A \le 2m.
    \]
    
    Therefore,
    \[
    m \ge \Big\lceil \frac{r(y) - r(x)}{2} \Big\rceil.
    \]

    Recalling that $d_{ID}(x,y) = 2m$, we obtain
    \[
    d_{ID}(x,y) \ge 2 \Big\lceil \frac{r(y) - r(x)}{2} \Big\rceil.
    \]

    If $r(x) \ge r(y)$, the same argument applies after swapping the roles of $x$ and $y$, yielding the bound in terms of $\lvert r(x) - r(y) \rvert$. Hence, in general,
    \[
        d_{ID}(x,y) \ge 2 \Big\lceil \frac{\lvert r(x) - r(y) \rvert}{2} \Big\rceil.
    \]

    If $\lvert r(x) - r(y) \rvert$ is odd, then
    \[
        2\Big\lceil \frac{\Delta}{2} \Big\rceil = \Delta + 1,
    \]
    where $\Delta = \lvert r(x) - r(y) \rvert$, completing the proof.
    \end{proof}

    We illustrate the tightness of the bound in Lemma~\ref{run_insdel_bound} with two minimal examples, considering both even and odd run-count differences.
    \begin{example}
    
       \noindent \textbf{Even difference.}  
        Let \(y = 0100\) and \(x = 0000\).  
        Here \(r(y) = 3\), \(r(x) = 1\), so \(|r(y)-r(x)| = 2\). The bound yields  
        \[
            d_{ID}(x,y) \ge 2\Big\lceil \frac{2}{2}\Big\rceil = 2.
        \]  
        The insdel distance between $x$ and $y$ is given by 
        \[
            d_{ID}(x,y) = 2(4-3) = 2.
        \]
        Hence, the bound is tight.

        \smallskip
        \noindent \textbf{Odd difference.}  
        Let \(y = 0001\) and \(x = 0000\).  
        Here \(r(y) = 2\), \(r(x) = 1\), so \(|r(y)-r(x)| = 1\). The bound yields  
        \[
            d_{ID}(x,y) \ge 2\Big\lceil \frac{1}{2}\Big\rceil = 2.
        \]  
        The insdel distance between $x$ and $y$ is given by 
        \[
            d_{ID}(x,y) = 2(4-3) = 2.
        \]
        Hence, the bound is tight in this case too.
    \end{example}

    \noindent 
        For an even number $d$ with $2 \le d \le 2k$, denote
        \[
        B_{ID}(x, d) \triangleq \{ v \in \mathbb{F}_2^k : d_{ID}(x, v) \le d \}
        \]
        as the insdel ball centered at $x$ of radius $d$. Then, we have the following bounds on $|B_{ID}(x, d)|$.

    \begin{lemma}\label{insdel_ball}
        Let $k, t$ be positive integers such that $k \ge t \ge 1$. Then, for any $x\in \mathbb{F}_2^k$, we have:
        $$ |B_{ID}(x, 2t)| \le \binom{r(x) + t - 1}{t}\Bigg( \sum_{i = 0}^{t}\binom{k}{i}\Bigg) \leq \Bigg(\frac{e^2(k+t)k}{t^2}\Bigg)^t,$$
        where $r(x)$ is the number-of-runs in $x$ and $e$ denotes Euler's number.
        Moreover, defining the worst-case insdel ball size as
        \[
        B^{\max}_{ID}(k,2t)
        \triangleq
        \max_{x \in \mathbb{F}_2^k} |B_{ID}(x,2t)|,
        \]
        it holds that
        \[
        B^{\max}_{ID}(k,2t)
        \le
        \left(\frac{e^2 (k+t)k}{t^2}\right)^t.
        \]
    \end{lemma}   
    \begin{proof}
        The first inequality in the upper bound follows from \cite[Theorem 4]{sala2013counting}. 
        For the second inequality, since $r(x)\le k$, we have,
        \begin{equation}
        \label{eq:ineq}
            \binom{r(x)+t-1}{t}\sum_{i=0}^{t}\binom{k}{i}
        \;\le\;
        \binom{k+t-1}{t}\sum_{i=0}^{t}\binom{k}{i}.
        \end{equation}

        \noindent
        We bound the two factors in Equation~\ref{eq:ineq} separately.\\
        Using $\binom{a}{b} \le \frac{a^b}{b!}$ together with $b! \ge (b/e)^b$, we obtain $\binom{a}{b} \le \left(\frac{ea}{b}\right)^b$. Applying this to $\binom{k+t-1}{t}$ yields 
        $$\binom{k+t-1}{t} \le \left(\frac{e(k+t)}{t}\right)^t.$$
        To handle the sum, note that $0 \leq i \leq t\leq k$, we have
        \[
            \Big(\frac{k}{t}\Big)^{t-i} \geq 1, 
        \]
        hence
        \[
            \binom{k}{i} \leq \binom{k}{i}\Big(\frac{k}{t}\Big)^{t-i}.
        \]
        Summing this inequality over $0\leq i\leq t$ gives
        \begin{align*}
           \sum_{i=0}^t\binom{k}{i} & \leq \sum_{i=0}^t \binom{k}{i}\Big(\frac{k}{t}\Big)^{t-i}\\
                                              & = \Big(\frac{k}{t}\Big)^{t} \sum_{i=0}^t \binom{k}{i} \Big(\frac{t}{k}\Big)^i \\
                                              & \leq \Big(\frac{k}{t}\Big)^{t} \sum_{i=0}^k \binom{k}{i}\Big(\frac{t}{k}\Big)^i \\
                                              & = \Big(\frac{k}{t}\Big)^{t} \Big( 1+\Big(\frac{t}{k}\Big)\Big)^k.
        \end{align*}
        Applying the elementary inequality $1 + u \leq e^u$ to $\Big(1+\Big(\frac{t}{k}\Big)\Big)$, we get
        \[
            \sum_{i=0}^t\binom{k}{i}  \leq  \Big(\frac{k}{t}\Big)^{t} e^t = \Big(\frac{ek}{t}\Big)^{t}.
        \]
        \noindent
        Combining these two estimates, we obtain
        \[
        \binom{k+t-1}{t}\sum_{i=0}^{t}\binom{k}{i}
        \;\le\;
        \left(\frac{e(k+t)}{t}\right)^t
        \left(\frac{ek}{t}\right)^t
        \;=\;
        \left(\frac{e^2 (k+t)k}{t^2}\right)^t.
        \] 
    \end{proof}

    \begin{remark}
        In the literature, an insdel ball is commonly referred to as a fixed-length Levenshtein ball. For a given center word, it is defined as the collection of all sequences of the same length that can be obtained from the center word by performing exactly $t$ deletions and exactly $t$ insertions. Equivalently, this set consists of all words whose insdel distance from the center word is at most $2t$ and whose length is preserved.
    \end{remark}
    
    \begin{lemma}[Singleton Bound]\cite{liu2022bounds}\label{singleton}
        For a binary insdel code $\mathcal{C}$ of length $n$ and minimum distance $d$ , one has
        \[
            |\mathcal{C}| \leq 2^{n - d/2 + 1}.
        \]
    \end{lemma}
    

 \subsection{Irregular-Distance Codes}
    In this section, we recall the notion of an irregular-distance code, which serves as the combinatorial bridge between function-correcting codes and classical error-correcting codes in the Hamming metric, together with the associated Plotkin-like bound. We also recall the definition of locally $(\lambda, \rho)_H$-function.
    
    \begin{definition}[Irregular distance code~\cite{10132545}]
    \label{def:irr-dis-code}
        Let $\textbf{D} \in \mathbb{N}_0^{M \times M}$. A set of binary codewords $\mathcal{P} = \{p_1, p_2, \ldots, p_M\}$ is called an irregular-distance code if there exists an ordering of the codewords such that $d_H(p_i, p_j) \ge [\textbf{D}]_{ij} \quad \text{for all } i,j \in [M].$ We define $N(\textbf{D})$ to be the smallest integer $r$ for which there exists a $\textbf{D}$-irregular-distance code ($\textbf{D}$-code) of length $r$.  
        If $[\textbf{D}]_{ij} = D$ for all $i \neq j$, we write $N(M, D).$
    \end{definition}

    \begin{lemma}[\cite{10132545}]\label{plotkin_FCC}
        For any distance matrix $\textbf{D} \in \mathbb{N}_0^{M \times M}$,
        \[
        N(\textbf{D}) \ge
        \begin{cases}
        \dfrac{4}{M^2} \displaystyle\sum_{i,j; \, i<j} [\textbf{D}]_{ij}, & \text{if $M$ is even}, \\[1.2em]
        \dfrac{4}{M^2 - 1} \displaystyle\sum_{i,j; \, i<j} [\textbf{D}]_{ij}, & \text{if $M$ is odd}.
        \end{cases}
        \]
    \end{lemma}

    \noindent
    For the next result, we recall the notions of Hamming ball, Hamming function ball, and locally bounded function in the Hamming metric. For $u \in \mathbb{F}_2^k$ and $\rho \in \mathbb{N}$, the Hamming ball of radius $\rho$ around $u$ is
    \[
    B_H(u, \rho) \triangleq 
    \{\, v \in \mathbb{F}_2^k \,:\, d_H(u, v) \leq \rho \,\}.
    \]
    \noindent
    For a function $f : \mathbb{F}_2^k \to \mathrm{Im}(f)$, the Hamming function ball of radius $\rho$ around $u$ is defined as
    \[
    B^f_H(u, \rho) \triangleq f\!\big(B_H(u, \rho)\big) 
    = \{\, f(v) \,:\, v \in B_H(u, \rho) \,\}.
    \]

    \begin{definition}[Locally $(\lambda, \rho)_H$-function~\cite{Rajput2025}]
    \label{def:local-H}
    A function $f : \mathbb{F}_2^k \to \mathrm{Im}(f)$ is said to be a 
    locally $(\lambda, \rho)_H$-function if 
    \[
    \big|B^f_H(u, \rho)\big| \leq \lambda 
    \qquad \text{for all } u \in \mathbb{F}_2^k.
    \]
    \end{definition}
    
    \begin{lemma}[\cite{Rajput2025}]\label{graph_color_function}
        Let $f : \mathbb{F}_2^k \rightarrow \mathrm{Im}(f)$ be a locally $(\lambda, \rho)_H$-function. 
        Assume that $\mathrm{Im}(f)$ is equipped with a total order $\prec$, and that for every $u \in \mathbb{F}_2^k$, the set $B_{\mathrm{H}}^f(u,\rho)$
        forms a contiguous block with respect to $\prec$.
        Then there exists a mapping
        \[
            \mathrm{Col}_f : \mathbb{F}_2^k \rightarrow [\lambda],
        \]
        such that for all $u,v \in \mathbb{F}_2^k$ satisfying $f(u) \neq f(v) \quad \text{and} \quad d_{H}(x,y) \leq \rho$, we have $\mathrm{Col}_f(u) \neq \mathrm{Col}_f(v)$.
    \end{lemma}

  
\section{Function-Correcting Codes for Insertion-Deletion Errors}
    
    In this section, we consider a function defined over the binary vector space $\mathbb{F}_2^k$, that is, $f: \mathbb{F}_2^k \rightarrow \mathrm{Im}(f)$, where the expressiveness of the function is given by $E = |\mathrm{Im}(f)|$. Let $x \in \mathbb{F}_2^k$ denote a binary vector on which the function $f$ is evaluated. The vector $x$ is transmitted over an asynchronous channel that may introduce at most $t$ insertions, at most $t$ deletions, or a combination of both. To enable correction of such errors, $x$ is first encoded using an encoding function $\psi(x) = (x, p(x))$, where $p(x) \in \mathbb{F}_2^r$ represents the redundancy added to $x$ prior to transmission. Based on this encoding framework, we define function-correcting deletion codes, function-correcting insertion codes, and function-correcting insertion-deletion (insdel) codes, which are designed to recover the function value $f(x)$ in the presence of deletion, insertion, or combined insertion-deletion errors, respectively.

    \begin{definition}[Function-Correcting Deletion Codes (FCDCs)]
        An encoding function $\psi(x): \mathbb{F}_2^k \rightarrow \mathbb{F}_2^{k+r}$ with $\psi(x) = (x, p(x))$, $x \in \mathbb{F}_2^k$ defines a function-correcting deletion code for $f: \mathbb{F}_2^k \rightarrow \mathrm{Im}(f)$ if for all $x$ and $y$ such that $f(x) \neq f(y)$, the following holds
        \[
           D_t(\psi(x)) \cap D_t(\psi(y)) = \emptyset.
        \]
    \end{definition}
    
    The aforementioned formulation guarantees that, even after up to $t$ deletions, the associated codewords for any two inputs $x$ and $y$ that result in different function values $f(x)$ and $f(y)$ remain recognizable. Consequently, any received subsequence resulting from at most $t$ deletions can be uniquely mapped back to a codeword, and hence to its corresponding function value $f(x)$.
    On the same line, function-correcting insertion codes can be defined as follows:
    
    \begin{definition}[Function-Correcting Insertion Codes (FCICs)]
        An encoding function $\psi(x): \mathbb{F}_2^k \rightarrow \mathbb{F}_2^{k+r}$ with $\psi(x) = (x, p(x))$, $x \in \mathbb{F}_2^k$ defines a function-correcting insertion code for $f: \mathbb{F}_2^k \rightarrow \mathrm{Im}(f)$ if for all $x$ and $y$ such that $f(x) \neq f(y)$, the following holds
        \[
           I_t(\psi(x)) \cap I_t(\psi(y)) = \emptyset.
        \]
    \end{definition}
    
    Function-correcting insdel codes, designed to recover function evaluations in the presence of both insertion and deletion errors, are defined as follows.
    
    \begin{definition}[Function-Correcting Insdel Codes (FCIDCs)]
    \label{def:fcidc}
        The encoding map $\psi: \mathbb{F}_2^k \rightarrow \mathbb{F}_2^{k+r}$ defined as $\psi(x) = (x, p(x))$ yields a function-correcting code for the function $f: \mathbb{F}_2^k \rightarrow \mathrm{Im}(f)$ in the insdel metric if for all $x,y \in \mathbb{F}_2^k$ such that $f(x) \neq f(y)$, the insdel distance $d_{ID}(\psi(x), \psi(y)) \geq 2t+2.$
    \end{definition}

    \begin{remark}
        The condition $d_{ID}(\psi(x),\psi(y))\ge 2t+2$ whenever $f(x)\ne f(y)$ guarantees that the function value $f(x)$ is recoverable from any output of a channel that introduces at most $t$ insdel errors. Let $\psi(x)$ be a codeword sent over such a channel, and let $w$ be the received word, so that $w$ is obtained from $\psi(x)$ by at most $t$ insertion and deletion operations and hence $d_{ID}(\psi(x),w)\le t$. The decoder outputs $f(\bar{x})$, where $\psi(\bar{x})$ is a codeword satisfying $d_{ID}(\psi(\bar{x}), w) \leq t$. Such a codeword exists because the transmitted codeword $\psi(x)$ is itself within distance $t$ of $w$. We claim $f(\bar{x})=f(x)$. By the triangle inequality for the insdel metric
        \[
        d_{ID}(\psi(x), \psi(\bar{x})) \leq d_{ID}(\psi(x), w) + d_{ID}(w, \psi(\bar{x})) \leq 2t.
        \]
        Were $f(x)\ne f(\bar{x})$, the FCIDC condition (Definition~\ref{def:fcidc}) would force $d_{ID}(\psi(x),\psi(\bar{x}))\ge 2t+2$, contradicting the above. Hence $f(\bar{x})=f(x)$, so the decoder recovers the correct function value even when the transmitted codeword is not uniquely identifiable.
    \end{remark}

    The next proposition establishes equivalence between the above-defined codes. 
    
    \begin{proposition}
        Let $\psi:\mathbb{F}_2^k\to\mathbb{F}_2^{n}$ be an encoding map with $n=k+r$, and fix $t\ge 1$. 
        The following conditions are equivalent for distinct inputs $x,y\in\mathbb{F}_2^k$ with $f(x)\neq f(y)$:
        \begin{enumerate}
          \item $D_t(\psi(x))\cap D_t(\psi(y))= \emptyset$.
          \item $I_t(\psi(x))\cap I_t(\psi(y))= \emptyset$.
          \item $d_{ID}(\psi(x),\psi(y))>2t$.
        \end{enumerate}
        In particular, a code that is function-correcting for up to $t$ deletions (insertions) is also function-correcting for up to $t$ insertion-deletions.
    \end{proposition}

    \begin{proof}
    Let $n=|\psi(x)|=|\psi(y)|$.
 
    \noindent (1) $\Rightarrow$ (3). If $D_t(\psi(x))\cap D_t(\psi(y))=\emptyset$, then there is no common subsequence of length at least $n-t$ for $\psi(x)$ and $\psi(y)$. Hence
    \[
    LCS(\psi(x),\psi(y)) < n-t.
    \]
    Therefore, \(d_{ID}(\psi(x),\psi(y)) = 2(n-LCS(\psi(x),\psi(y))) > 2t\).
    
    \smallskip
    \noindent (3) $\Rightarrow$ (1). If $d_{ID}(\psi(x),\psi(y))>2t$, then 
    \(LCS(\psi(x),\psi(y))<n-t\). Hence, there cannot exist a common subsequence of length \(\ge n-t\). Therefore, we get 
    
    $$D_t(\psi(x))\cap D_t(\psi(y))=\emptyset.$$

    \smallskip
    \noindent (1) $\Rightarrow$ (2). We proceed by contradiction. Suppose that there exists a word \(w\) with \(w\in I_t(\psi(x))\cap I_t(\psi(y))\). By the definition of $I_t(\cdot)$, the word $w$ is obtained from each of $\psi(x)$ and $\psi(y)$ by inserting exactly $t$ symbols. Hence $|w| = n + t$ and both $\psi(x)$ and $\psi(y)$ are subsequences of $w$. Therefore, each of $\psi(x)$ and $\psi(y)$ can be obtained from $w$ by deleting exactly $t$ of its $n+t$ symbols, so at most $2t$ positions of $w$ are deleted in forming the two sequences. Consequently, at least $(n+t) - 2t = n - t$ positions of $w$ are retained in both, and the symbols in these positions form a common subsequence of $\psi(x)$ and $\psi(y)$. Therefore
    \[
    LCS(\psi(x), \psi(y)) \geq n - t.
    \]
    Let $z$ be a common subsequence of length exactly $n - t$. Then $z$ is obtained from each of $\psi(x)$ and $\psi(y)$ by deleting exactly $t$ symbols, so $z \in D_t(\psi(x)) \cap D_t(\psi(y))$, contradicting $(1)$. Hence $I_t(\psi(x)) \cap I_t(\psi(y)) = \emptyset$.

    \smallskip
    \noindent (2) $\Rightarrow$ (3) 
    Suppose, for contradiction, that 
    \[
    d_{ID}(\psi(x), \psi(y))  \;\leq\; 2t,
    \]
    so there exists a sequence of at most $2t$ operations (insertions and deletions) transforming $\psi(x)$ into $\psi(y)$. Since $|\psi(x)| = |\psi(y)| = n$, the number of insertions must equal the number of deletions in any such transformation. We denote this common count by $i$, where $i \leq t$.

    We construct an intermediate word $w$ by performing all $i$ insertions on $\psi(x)$ first, before any deletions. Then
    \begin{itemize}
    \item $\psi(x)$ is a subsequence of $w$, since $w$ is obtained from 
    $\psi(x)$ by $i$ insertions;
    \item $\psi(y)$ is a subsequence of $w$, since $w$ is reduced to $\psi(y)$ 
    by $i$ deletions;
    \item $|w| = n + i \leq n + t$.
    \end{itemize}

    If $|w| < n + t$, extend $w$ to a word $w'$ of length exactly $n + t$ by inserting $n + t - |w|$ arbitrary symbols at any positions; inserting symbols preserves the subsequence relation, so $\psi(x)$ and $\psi(y)$ remain subsequences of $w'$. Since $|w'| - |\psi(x)| = |w'| - |\psi(y)| = t$, the word $w'$ is obtained from each of $\psi(x)$ and $\psi(y)$ by inserting exactly $t$ symbols, so $w' \in I_t(\psi(x)) \cap I_t(\psi(y))$. This contradicts the hypothesis $I_t(\psi(x)) \cap I_t(\psi(y)) = \emptyset$.

    \smallskip
    \noindent (2) $\Rightarrow$ (1) can be deduced from (2)\(\Rightarrow\)(3)\(\Rightarrow\)(1).
    \end{proof}  
    Given the equivalence among the three code formulations discussed above, we restrict our attention, in the rest of this paper, to {function-correcting insertion-deletion codes} (FCIDCs).
    
    Next, we define the optimal redundancy of an FCIDC for a function $f$. This is a key parameter that plays a central role in the study of FCIDCs.
    
    \begin{definition}[Optimal Redundancy]
    The optimal redundancy of a function-correcting insdel code for $f$ is the smallest integer $r$ such that there exists an encoding map $\psi: \mathbb{F}_2^k \to \mathbb{F}_2^{k+r}$ defining an FCIDC for $f$.
        The optimal redundancy is denoted by \( r^{f}_{ID}(k, t) \).
    \end{definition}
    

    \subsection{Irregular Insdel-Distance Codes and Their Connection to FCIDCs}

    In this section, we define irregular insdel-distance codes and establish their relationship with FCIDCs. Leveraging this connection, we derive several general results about FCIDCs and obtain both lower and upper bounds on their optimal redundancy.

    We first define insdel-distance matrices (Definition~\ref{insdel-matrices}) for a function $f$, followed by irregular insdel-distance codes (Definition~\ref{IIDC}) in which the insdel-distance between each pair of codewords should satisfy individual distance constraints.
    
    \begin{definition}[Insdel-Distance Matrices]
    \label{insdel-matrices} 
    Let $M,t \in \mathbb{N}$. Consider $M$ binary vectors ${x_1}, \ldots, {x_{M}} \in {\mathbb{F}_2^k}$. Then,  ${\textbf{I}_f^{(1)}}(t, {x_1}, \ldots , {x_{M}})$ and
    ${\textbf{I}_f^{(2)}}(t, x_1, \ldots, x_{M})$  are $M \times M$  insdel-distance matrices corresponding to function $f$ with entries as follows:
   
        \begin{equation*}    
        [\mathbf{I}_f^{(1)}(t, x_1,\dots,x_M)]_{ij} =
        \begin{cases}
        [2t + 2 - d_{ID}(x_i,x_j)]^{+}, & \text{if } f(x_i)\neq f(x_j),\\
        0, & \text{otherwise}.
        \end{cases}
    \end{equation*}

    \noindent {and}

    \begin{equation*}
    [\mathbf{I}_f^{(2)}(t, x_1,\dots,x_M)]_{ij} =
    \begin{cases}
    [2t + 2 + 2k - d_{ID}(x_i,x_j)]^{+}, & \text{if } f(x_i)\neq f(x_j),\\
    0, & \text{otherwise}.
    \end{cases}
    \end{equation*}   
    \end{definition}

    \begin{remark}
    \label{rmk:2k-term}
    Unlike the Hamming metric, the insdel metric admits cross-block alignment between concatenated words (Lemma~\ref{insdel_inequality}), which may reduce the insdel distance of a concatenation by up to $2\min\{k, r\}$. The additional $2k$ term in the type 2 matrix $\bm{I}^{(2)}_f$ is therefore introduced to offset this loss when $r \geq k$. This compensation is indispensable and the type 1 condition alone does not suffice to guarantee a function-correcting insdel code. We illustrate this necessity in Example~\ref{ex:2k-essential}.
    \end{remark}
    
    An illustrative example corresponding to each of the two types of matrices is presented next. 
    
    \begin{example}\label{eg_dis_matrix}
        Let $f:\{0,1\}^2\to\{0,1\}$ be given by
        \[
        f(10)=f(01)=0 \text{ and } f(00) =  f(11)=1.
        \]
        Fix the input ordering
        \[
        (x_1,x_2,x_3,x_4)=(00,01,10,11).
        \]
        For $t=1$, the matrices $\textbf{I}_f^{(1)}(t,x_1,x_2,x_3,x_4)$ and $\textbf{I}_f^{(2)}(t,x_1,x_2,x_3,x_4)$ are
        \[
         \textbf{I}_f^{(1)}(t, x_1,x_2,x_3,x_4)=
         \begin{pmatrix}
          0 & 2 & 2 & 0\\
          2 & 0 & 0 & 2\\
          2 & 0 & 0 & 2\\
          0 & 2 & 2 & 0
         \end{pmatrix},
         \]
         \[
         \textbf{I}_f^{(2)}(t, x_1, x_2, x_3, x_4)=
         \begin{pmatrix}
          0 & 6 & 6 & 0\\
          6 & 0 & 0 & 6\\
          6 & 0 & 0 & 6\\
          0 & 6 & 6 & 0
         \end{pmatrix}.
        \]       
    \end{example}


    \begin{definition}[Insdel distance-constraint matrix]
    \label{def:dist-constraint-matrix}
    An insdel distance-constraint matrix of order $M$ is a matrix $\mathbf{I} \in \mathbb{N}_0^{M \times M}$ whose entry $[\mathbf{I}]_{ij}$ specifies a required lower bound on the insdel distance between the $i$-th and $j$-th codewords of a length-$r$ binary code. The matrices $\mathbf{I}^{(1)}_f$ and $\mathbf{I}^{(2)}_f$ of Definition~\ref{insdel-matrices}, and the function distance matrices of Definition~\ref{def:function-dis-matrices}, are instances of such matrices.
    \end{definition}

    Let $\mathcal{P} = \{p_1, p_2, \dots, p_M\} \subseteq \mathbb{F}_2^r$ be a code of size $M$ and length $r$. The unconventional choice of using the code-block length $r$ is motivated by its relationship to the redundancy of FCIDCs, which is discussed later in this section.
    
    \begin{definition}[Irregular insdel-distance codes]
    \label{IIDC}
        Let $\mathbf{I}$ be an insdel distance-constraint matrix of order $M$ (Definition~\ref{def:dist-constraint-matrix}), and let $K \in \mathbb{N}$.
        Then,
        \begin{itemize}
          \item $\mathcal{P}$ is an irregular insdel-distance code of type~ $1$ for matrix $\mathbf{I}$ if there exists an ordering of $\mathcal{P}$ such that
          \[
            d_{ID}(p_i,p_j) \ge [\bm{I}]_{i,j} \quad \text{for all } 1\le i,j\le M.
          \]
          \item $\mathcal{P}$ is an irregular insdel-distance code of type $2$ corresponding to matrix $\mathbf{I}$ if it is an irregular insdel-distance code of type~ $1$ and, in addition, its codeword length satisfies $r \ge K$.
        \end{itemize}
    \end{definition}

    Next, we define the shortest achievable length for both types of irregular insdel-distance codes.
    
    \begin{definition}
       \begin{align*}
            &N^{(1)}_{ID}(\bm{I})
                := \min\{\, r : \exists\ \text{type 1 irregular insdel-distance code of length } r \,\},\\[4pt]
            &N^{(2)}_{ID}(\bm{I};K)
                := \min\{\, r \ge K : \exists\ \text{type 2 irregular insdel-distance code of length } r \,\}.
        \end{align*}
        When $[\bm{I}]_{i,j} = I$ for all $i \neq j$, where $I \in \mathbb{N}$, we write $N^{(1)}_{ID}(\bm{I})$ and $N^{(2)}_{{ID}}(\bm{I};K)$ as $N^{(1)}_{ID}(M,I)$ and $N^{(2)}_{{ID}}(M,I;K)$,  respectively.
    \end{definition}

    \begin{example}\label{eg_irregular_code}
        Let $p_1 = (0) , p_2 = (1), p_3 = (1)$ and $p_4 = (0)$. Then, $\{p_1, p_2, p_3, p_4\}$ is an irregular insdel-distance code of type $1$ corresponding to matrix $\textbf{I}_f^{(1)}(t, x_1,x_2,x_3,x_4)$ of Example~\ref{eg_dis_matrix}. Clearly,  $N^{(1)}_{{ID}}(\textbf{I}_f^{(1)}(t, x_1,x_2,x_3,x_4)) = 1.$

        \smallskip
        \noindent Now, let $K = 2$. Then $p_1 = (000) , p_2 = (111), p_3 = (111)$ and $p_4 = (000)$ is an irregular insdel-distance code of type $2$ corresponding to matrix $\textbf{I}_f^{(2)}(t, x_1,x_2,x_3,x_4)$ of Example~\ref{eg_dis_matrix}. In this case $N^{(2)}_{{ID}}(\textbf{I}_f^{(2)}(t, x_1,x_2,x_3,x_4);2) = 3.$
    \end{example}
    
    The results in the next two lemmas play a crucial role in establishing the relationship between the optimal redundancy of FCIDCs and the lengths of irregular insdel-distance codes.
     
    \begin{lemma}
        Let $x = (x_1, x_2) \in \mathbb{F}_2^{k+r}$ and $y = (y_1, y_2) \in \mathbb{F}_2^{k+r}$ where $x_1, y_1 \in \mathbb{F}_2^k$ and $x_2, y_2 \in  \mathbb{F}_2^r$, then, 
        \begin{equation}\label{lcs_inequality}  
        \begin{multlined}
            LCS(x_1, y_1) + LCS(x_2, y_2) \leq LCS(x,y) \leq LCS(x_1, y_1) + LCS(x_2, y_2) + \min\{k,r\}
        \end{multlined}           
        \end{equation}    
    \end{lemma}    
    
    \begin{proof}
        Let $s_1$ be the longest common subsequence of $x_1$ and $y_1$, and $s_2$ be the longest common subsequence of $x_2$ and $y_2$; then $s_1s_2$ is a common subsequence of $x$ and $y$. Therefore,
        \begin{equation*}
            LCS(x,y) \geq LCS(x_1, y_1) + LCS(x_2, y_2)
        \end{equation*}
        For the right-hand inequality, let $s$ be the longest common subsequence of $x$ and $y$. For each symbol in $s$, consider the positions to which it is matched in both $x$ and $y$. This induces a partition of the symbols of $s$ into four classes indexed by $(i,j), \quad i,j \in \{1,2\}$, where a symbol is of type $(i,j)$ if it is matched to $x_i$ in  $x$ and to $y_j$ in $y$. Let $n_{ij}$ denote the number of symbols of type $(i,j)$ in $s$. The following four cases can be considered:
        
        \noindent\textbf{Case 1:} If the symbols are of type $(1,1)$, then they form a common subsequence of $x_1$ and $y_1$. Therefore,
        \[
            n_{11} \leq LCS(x_1,y_1).
        \]
        \textbf{Case 2:} If the symbols are of type $(2,2)$, then they form a common subsequence of $x_2$ and $y_2$. Therefore,
        \[
            n_{22} \leq LCS(x_2,y_2).
        \]
        \textbf{Case 3:} If the symbols are of type $(1,2)$, then they form a common subsequence of $x_1$ and $y_2$. Therefore,
        \[
            n_{12} \leq \min\{k,r\}.
        \]
        \textbf{Case 4:} If the symbols are of type $(2,1)$, then they form a common subsequence of $x_2$ and $y_1$. Therefore,
        \[
            n_{21} \leq \min \{k,r\}.
        \]
        We claim $n_{12}\cdot n_{21} = 0,$ i.e., symbols of type $(1,2)$ and $(2,1)$ cannot appear together in $s$.
        On the contrary assume that there exists symbols $s_i$ and $s_j$ of type $(1,2)$ and $(2,1)$ of $s$, if $s_i$ precedes $s_j$, then the ordering is respected in $x$ but violated in $y$. Similarly, if $s_j$ precedes $s_i$, then the ordering is respected in $y$ but violated in $x$.
        
        \noindent Hence, the claim follows and we have $n_{12} + n_{21} = \max\{n_{12}, n_{21}\} \leq \min \{k,r\}.$
    \end{proof}
    
    \begin{lemma}\label{insdel_inequality}
          Let $x = (x_1, x_2) \in \mathbb{F}_2^{k+r}$ and $y = (y_1, y_2) \in \mathbb{F}_2^{k+r}$ where $x_1, y_1 \in \mathbb{F}_2^k$ and $x_2, y_2 \in  \mathbb{F}_2^r$, then,
          
        \begin{equation}\label{insdel_partition}
              \begin{multlined}
                d_{ID}(x_1, y_1) + d_{ID}(x_2, y_2) - 2 \cdot \min\{k,r\}\leq d_{ID}(x,y) \leq d_{ID}(x_1, y_1) + d_{ID}(x_2, y_2) 
              \end{multlined}    
        \end{equation}           
    \end{lemma}
    
    \begin{proof}
        From the definition of the insdel-distance, we have
        \begin{align*}
            &d_{ID}(x, y) = 2(k+r) - 2LCS(x,y)\\
            &d_{ID}(x_1, y_1) = 2k - 2LCS(x_1,y_1)\\
            &d_{ID}(x_2,y_2) = 2r - 2LCS(x_2, y_2)
        \end{align*}
        Multiplying equation~\eqref{lcs_inequality} by $-2$, we get:
        \begin{align*}
            -2(LCS(x_1, y_1) + LCS(x_2, y_2) + \min\{k,r\})&\leq -2 \cdot LCS(x,y)\\
                                                         &\leq  -2(LCS(x_1, y_1) + LCS(x_2, y_2))
        \end{align*}
        Adding $2(k+r)$ to each part in the above inequality, we get
        
        \begin{align*}
            2(k+r) - 2(LCS(x_1, y_1) + LCS(x_2, y_2) + \min\{k,r\})& \leq  2(k+r) -2 \cdot LCS(x,y)\\
                                                                 &\leq  2(k+r) -2(LCS(x_1, y_1) + LCS(x_2, y_2))
        \end{align*}

        \begin{align*}
            2k - 2 \cdot LCS(x_1, y_1) + 2r - 2LCS(x_2, y_2) - 2\min\{k,r\} &\leq  2(k+r) - 2LCS(x,y)\\
                                                                               &\leq 2k - 2 \cdot LCS(x_1, y_1) + 2r - 2LCS(x_2, y_2)
        \end{align*}
        Therefore,
        \begin{align*}
            d_{ID}(x_1, y_1) + d_{ID}(x_2, y_2) - 2 \cdot \min\{k,r\} \leq d_{ID}(x,y)
                                                                     \leq d_{ID}(x_1, y_1) + d_{ID}(x_2, y_2)
        \end{align*}        
    \end{proof}

    \begin{remark}
    \label{rmk:redundancy-regime}
    Lemma~\ref{insdel_inequality} establishes that cross-block longest common subsequence effects can reduce the insdel distance between concatenated codewords by as much as $2\min\{k, r\}$. Consequently, for any systematic encoding $\psi(x) = (x, p(x))$ with redundancy length $r$, the function-correcting insdel code requirement ($d_{ID}(\psi(x_i), \psi(x_j)) \geq 2t + 2, \text{ whenever } f(x_i) \neq f(x_j))$ is guaranteed, via the lower bound in Lemma~\ref{insdel_inequality}, by the following sufficient condition on the redundancy vectors:
    \begin{equation}
    \label{eq:min{k,r}}
        d_{ID}(p_i, p_j) \;\geq\; 2t + 2 - 
    d_{ID}(x_i, x_j) + 2\min\{k, r\}. 
    \end{equation}

    \noindent Consider the following cases based on the relative sizes of $k$ and $r$.

    \smallskip
    \noindent\textbf{Case 1:} Let $r \geq k$. In this case $\min\{k, r\} = k$, and equation~\eqref{eq:min{k,r}} reduces to
    \[
    d_{ID}(p_i, p_j) \;\geq\; 2t + 2 + 2k - 
    d_{ID}(x_i, x_j),
    \]
    which is precisely the type~2 distance condition encoded in the matrix $I^{(2)}_f$. The $+2k$ term in the type~2 matrix exactly compensates for the worst-case cross-block LCS loss.

    \smallskip
    \noindent\textbf{Case 2: }Let $r < k$. Here $\min\{k, r\} = r$, and equation~\eqref{eq:min{k,r}} becomes
    \[
    d_{ID}(p_i, p_j) \;\geq\; 2t + 2 - 
    d_{ID}(x_i, x_j) + 2r.
    \]
    However, the trivial upper bound $d_{ID}(p_i, p_j) \leq 2r$ (since $p_i, p_j \in \mathbb{F}_2^{r}$), together with equation~\eqref{eq:min{k,r}}, forces
    \[
    2r \;\geq\; 2t + 2 - d_{ID}(x_i, x_j) + 2r,
    \qquad \text{that is,} \qquad 
    d_{ID}(x_i, x_j) \;\geq\; 2t + 2.
    \]
    Thus, in the regime $r < k$, the message vectors $x_i$ and $x_j$ must themselves already satisfy the full insdel distance constraint of a classical $t$-insdel error-correcting code. The redundancy contributes nothing toward relaxing the message-side distance requirement. 

    \smallskip
    This motivates our focus on the type~2 framework with $r \geq k$. Accordingly, in Definition~\ref{IIDC} the type~2 irregular insdel-distance code is required to satisfy $r \geq K$, and Theorem~\ref{Theorem_opt_irr} sets $K = k$.
    \end{remark}
    
    The following example shows that the lower bound given in Lemma~\ref{insdel_inequality} is tight.
    
    \begin{example}
        Let $x = (0000, 10111) \in \mathbb{F}_2^{9}$ and $y = (1111,00000) \in \mathbb{F}_2^{9}$. Then,
        $d_{ID}(0000,1111) =~8,\\ \quad d_{ID}(10111,00000)= 8 \text{ and } d_{ID}(000010111, 111100000) = 8$. 
        Hence, $d_{ID}(000010111, 111100000) = d_{ID}(0000,1111) + d_{ID}(10111,00000) - 2\cdot4$.
    \end{example}

    The lower bound of Lemma~\ref{insdel_inequality} includes an extra term $\min\{k,r\}$, resulting from cross-block alignments. The next result shows that cross-block alignments can never reduce the concatenation distance below that of the second block alone. 

    \begin{lemma}
    \label{lem:block-distance}
    Let $x,y\in\mathbb{F}_2^{k}$ and $x',y'\in\mathbb{F}_2^{r}$. Then
    \[
    LCS\big((x,x'),(y,y')\big)
    \;\le\;
    \min\big\{\, k+LCS(x',y'),\;\;
                 r+LCS(x,y) \,\big\},
    \]
    and consequently
    \[
    d_{ID}\big((x,x'),(y,y')\big)
    \;\ge\;
    \max\big\{\, d_{ID}(x,y),\;\; d_{ID}(x',y') \,\big\}.
    \]
    \end{lemma}

    \begin{proof}
    Write $X=(x,x')$ and $Y=(y,y')$, both of length $k+r$, and let $s$ be a longest common subsequence. As in the proof of Lemma~\ref{insdel_inequality}, classify each matched symbol of $s$ by the pair $(i,j)$ recording whether its position in $X$ and in $Y$ falls in the length-$k$ prefix ($1$) or the length-$r$ suffix ($2$). Let $n_{ij}$ be the corresponding counts, so $LCS(X,Y)=n_{11}+n_{12}+n_{21}+n_{22}$, and recall that $n_{12}\,n_{21}=0$.

    The type-$(1,1)$ symbols match positions of $x$ to positions of $y$ in increasing order, hence form a common subsequence of $x$ and $y$, likewise the type-$(2,2)$ symbols form a common subsequence of $x'$ and $y'$. Therefore
    \[
    n_{11}\le LCS(x,y),
    \qquad
    n_{22}\le LCS(x',y').
    \]
    Counting distinct positions used in each block gives $n_{11}+n_{12}\le k$ and $n_{11}+n_{21}\le k$, and $n_{21}+n_{22}\le r$ and $n_{12}+n_{22}\le r$. Since $n_{12}n_{21}=0$, whichever cross term vanishes yields
    \[
    n_{11}+n_{12}+n_{21}\le k
    \qquad\text{and}\qquad
    n_{12}+n_{21}+n_{22}\le r .
    \]
    Combining the two inequalities above, we get,
    \[
    LCS(X,Y)=(n_{11}+n_{12}+n_{21})+n_{22}\le k+LCS(x',y'),
    \]
    \[
    LCS(X,Y)=n_{11}+(n_{12}+n_{21}+n_{22})\le r+LCS(x,y).
    \]
    Therefore, the insdel distance between $X$ and $Y$ is bounded from below 
    \[
    d_{ID}(X,Y)=2(k+r)-2LCS(x,y)
    \ge\max\{\,2k-2LCS(x,y),\;2r-2LCS(x',y')\,\}
    =\max\{\,d_{ID}(x,y),\,d_{ID}(x',y')\,\}. 
    \]
\end{proof}

    \begin{corollary}
    \label{cor:append-construction}
    Let $t \in \mathbb{N}$ and $f:\mathbb{F}_2^{k}\to \mathrm{Im}(f)$ with $E=|\mathrm{Im}(f)|$. Then
    \[
        r^{f}_{ID}(k,t) \;\le\; N^{(1)}_{ID}\big(E,\, 2t+2\big).
    \]
    where $N^{(1)}_{ID}\big(E,\, 2t+2\big)$ is the minimum length of a binary error-correcting insdel code with $E$ codewords and minimum distance $2t+2$.
    \end{corollary}

    \begin{proof}
    Fix an injective labelling $\mathrm{Im}(f)=\{f_0,\dots,f_{E-1}\}$ with index map $\iota(f_j)=j$, and let $\mathcal{D}=\{D_0,\dots,D_{E-1}\}\subseteq\mathbb{F}_2^{\ell}$ be a code with $E$ codewords, minimum insdel distance $d_{ID}(\mathcal{D})\ge 2t+2$, and minimum length $\ell=N^{(1)}_{ID}(E,2t+2)$. Define $\psi(u)=(u,\,D_{\iota(f(u))})$. For $u,u'\in\mathbb{F}_2^{k}$ with $f(u)\ne f(u')$, the codewords $D_{\iota(f(u))}$ and $D_{\iota(f(u'))}$ are distinct, so $d_{ID}(D_{\iota(f(u))},D_{\iota(f(u'))})\ge 2t+2$, and Lemma~\ref{lem:block-distance} gives $d_{ID}(\psi(u),\psi(u'))\ge 2t+2$. Hence $\psi$ is a $t$-FCIDC with redundancy $N^{(1)}_{ID}(E,2t+2)$.
    \end{proof}

    The following theorem establishes upper and lower bounds on the optimal redundancy of FCIDCs constructed for generic functions. Moreover, it reduces the problem of determining \( r^{f}_{ID}(k,t) \) to determining the quantities \( N_{ID}^{(1)}\left( \boldsymbol{I}^{(1)}_{f}(t, x_{1}, \ldots, x_{2^k})\right) \) and \( N_{ID}^{(2)}\!\left(\boldsymbol{I}^{(2)}_{f}(t, x_{1}, \ldots, x_{2^k})\right) \).

    \begin{theorem}\label{Theorem_opt_irr}  Let $f: \mathbb{F}_2^k\rightarrow \mathrm{Im}(f)$ be any function and let $\{x_1, x_2, \ldots, x_{2^k}\} = \mathbb{F}_2^k$.  
    
        \[
        N^{(1)}_{ID}(\boldsymbol{I}_f^{(1)}(t,x_1,x_2, \ldots ,x_{2^k})) \leq r_{ID}^f(k,t) \leq N^{(2)}_{ID}(\boldsymbol{I}_f^{(2)}(t,x_1,x_2, \ldots,x_{2^k});k).
        \]   
        
    \end{theorem}
    \begin{proof} We establish the theorem by considering the following cases.
    
        \medskip\noindent\textbf{Case 1: }(Constant functions)
        If \( f \) is a constant function, then \( N^{(1)}_{ID}(\boldsymbol{I}_f^{(1)}(t, x_1, \ldots,x_{2^k}))  = 0 \) and \( N^{(2)}_{ID}(\boldsymbol{I}_f^{(2)}(t, x_1, \ldots, x_{2^k});k)  = k\). Consequently, the desired condition holds trivially, establishing the theorem in this case. Specifically, we have:  
            \[
            N^{(1)}_{ID}(\boldsymbol{I}_f^{(1)}(t)) = r_{ID}^f(k,t) \leq N^{(2)}_{ID}(\boldsymbol{I}_f^{(2)}(t)).
            \]
        \medskip\noindent\textbf{Case 2: }(Non-constant functions)
        To prove that \( N^{(1)}_{ID}(\boldsymbol{I}_f^{(1)}(t, x_1,x_2, \ldots,x_{2^k})) \leq r_{ID}^f(k,t) \) for a non-constant function \( f \), consider a function-correcting insdel code for $f$ defined by an encoding function
        \[
            \psi: \mathbb{F}_2^k \to \mathbb{F}_2^{k+r}, \quad x_i \mapsto (x_i, {p_i}),
         \]
        where the redundancy \( r \) is optimal, i.e.,  \( r = r_{ID}^f(k,t) \). On the contrary, suppose that \\ \( N^{(1)}_{ID}(\boldsymbol{I}_f^{(1)}(t, x_1,x_2, \ldots,x_{2^k})) > r_{ID}^f(k,t) \). This implies the existence of distinct indices \( i,j \in \{1, \ldots, 2^k\} \) such that \( f({x_i}) \neq f({x_j}) \) and  
        \[
        d_{ID}({p_i},{p_j}) < 2t + 2 - d_{ID}({x_i},{x_j}).
        \]
        Consequently,  from equation~\eqref{insdel_partition} we obtain  
        \[
        d_{ID}(\psi({x_i}), \psi({x_j})) \leq d_{ID}({x_i}, {x_j}) + d_{ID}({p_i}, {p_j})  < 2t + 2.
        \]
        This contradicts that $\psi$ defines a function-correcting insdel code. Hence, 
        \[
        N^{(1)}_{ID}(\boldsymbol{I}_f^{(1)}(t, x_1,x_2, \ldots,x_{2^k})) \leq r_{ID}^f(k,t).
        \]
        \begin{sloppypar}
        Next, we establish the reverse inequality \( r_{ID}^f(k,t) \leq N^{(2)}_{ID}(\boldsymbol{I}_f^{(2)}(t, x_1,x_2, \ldots,x_{2^k});k) \). Let $\mathcal{P} = \{{p_1}, \ldots, {p_{2^k}}\}$ be an \(\boldsymbol{I}_f^{(2)}(t, x_1,x_2, \ldots,x_{2^k}) \) irregular insdel-distance code of type 2 and length \( r = N^{(2)}_{ID} (\boldsymbol{I}_f^{(2)}(t,  x_1,x_2, \ldots,x_{2^k});k) \), and define the encoding function  
        \end{sloppypar}
        \[
        \psi: \mathbb{F}_2^k \to \mathbb{F}_2^{k+r}, \quad {x_i} \mapsto ({x_i}, {p_i}).
        \]  
        For every pair \( i, j \in \{1, \ldots, 2^k\} \) with \( f({x_i}) \neq f({x_j}) \), we have  
        \begin{align*}
        d_{ID}(\psi({x_i}), \psi({x_j})) &\geq d_{ID}({x_i}, {x_j}) + d_{ID}({p_i}, {p_j}) \\& \quad -2\cdot \min\{k, N^{(2)}_{ID} (\boldsymbol{I}_f^{(2)}(t,  x_1, \ldots,x_{2^k});k) \}.\\
        &\geq d_{ID}({x_i}, {x_j}) + d_{ID}({p_i}, {p_j}) - 2k\\
        & \geq d_{ID}({x_i}, {x_j}) + 2t + 2 + 2k - d_{ID}({x_i}, {x_j})\\&\qquad -2k\\
        &= 2t + 2.
        \end{align*}
        
        Thus, \( \psi \) defines a function-correcting insdel code for \( f \) with redundancy \( r = N^{(2)}_{ID}(\boldsymbol{I}_f^{(2)}(t,  x_1,x_2, \ldots,x_{2^k});k) \), yielding  
        \[
        r^f_{ID}(k, t) \leq N^{(2)}_{ID}(\boldsymbol{I}_f^{(2)}(t,  x_1,x_2, \ldots,x_{2^k});k).
        \]        
        This completes the proof.  
    \end{proof}

    \begin{example}
        With the help of Example~\ref{eg_irregular_code} and Theorem~\ref{Theorem_opt_irr} we can bound the optimal redundancy of function $f$ defined in Example~\ref{eg_dis_matrix} as follows:
        \[
            1 \leq r_{ID}^f(2,1) \leq 3.
        \]
    \end{example}

    \begin{remark}
    Theorem~\ref{Theorem_opt_irr} derives the lower bound from the type~1 matrix $I^{(1)}_f$ and the upper bound from the type~2 matrix $I^{(2)}_f$. One might ask whether the simpler type~1 condition already suffices to guarantee a function-correcting insdel code, in which case the upper bound could also be obtained from $I^{(1)}_f$. The following example demonstrates that this is not the case: the $+2k$ compensation term in $I^{(2)}_f$ is indispensable.
    \end{remark}

    \begin{example}
    \label{ex:2k-essential}
    Let $k = 4$ and $t = 4$, and consider the two message vectors $x_1 = 0000$ and $x_2 = 1111$ in $\mathbb{F}_2^4$, with $f(x_1) \neq f(x_2)$. Encode each systematically as $\psi(x_i) = (x_i, p(x_i))$, where the redundancy vectors satisfy $p(x_i) \in \mathbb{F}_2^{6}$, and require that $\{p(x_1), p(x_2)\}$ form an irregular insdel-distance code of type~1 for $I^{(1)}_f$.

    Since $d_{ID}(x_1, x_2) = d_{ID}(0000, 1111) = 8$, the type~1 condition $d_{ID}(p(x_1), p(x_2)) \geq [I^{(1)}_f(t, x_1, x_2)]_{12}$ reduces to
    \[
    d_{ID}\big(p(x_1), p(x_2)\big) 
    \geq \big[\,2t + 2 - d_{ID}(x_1, x_2)\,\big]^{+} 
    = [\,10 - 8\,]^{+} = 2.
    \]
    The choice $p(x_1) = 101110$ and $p(x_2) = 000000$ satisfies this requirement, since $d_{ID}(p(x_1), p(x_2)) = 8 \geq 2$. However, for the resulting codewords $\psi(x_1) = 0000\,101110$ and $\psi(x_2) = 1111\,000000$, the longest common subsequence is the contiguous string of six zeros, yielding $\mathrm{LCS}(\psi(x_1), \psi(x_2)) = 6$ and consequently
    \[
    d_{ID}\big(\psi(x_1), \psi(x_2)\big) 
    = 2(10 - 6) = 8 < 2t + 2.
    \]
    Thus, this type~1-valid encoding fails to correct $t = 4$ insdel errors. In contrast, the type~2 condition imposes the stricter 
    requirement
    \[
    d_{ID}\big(p(x_1), p(x_2)\big) 
    \geq \big[\,2t + 2 + 2k - d_{ID}(x_1, x_2)\,\big]^{+} 
    = 10,
    \]
    which excludes the above choice of redundancy vectors. The failure stems precisely from the cross-block LCS loss formalized in Lemma~\ref{insdel_inequality}.
    \end{example}

 \subsection{Simplified Redundancy Lower Bounds} 
    Using a smaller set of  information vectors, one can obtain a lower bound on the redundancy of FCIDCs as follows:
    \begin{corollary}\label{lower_bound_red}
        Consider a collection of $M$ distinct binary words $x_1, x_2, \ldots, x_M$ of length $k$. Then, for a function $f$, the optimal redundancy of an FCIDC is lower bounded as follows:
        $$ r^f_{ID}(k, t) \geq N^{(1)}_{ID}(\boldsymbol{I}_f^{(1)}(t, x_1,x_2, \ldots,x_{M})).$$
        For a non-constant function $f$, 
        $$ r^f_{ID}(k, t) \geq N^{(1)}_{ID}(2, 2t) = t. $$
    \end{corollary}

    \begin{proof}
         Let $\mathcal{P}=\{p_1, p_2, \ldots, p_{2^k}\}$ be a $\boldsymbol{I}_f^{(1)}(t, x_1,x_2, \ldots,x_{2^k})$-code of length $N^{(1)}_{ID}(\boldsymbol{I}_f^{(1)}(t, x_1,x_2, \ldots,x_{2^k}))$; then $\mathcal{P}=\{p_1, p_2, \ldots, p_{M}\}$ is a $\boldsymbol{I}_f^{(1)}(t, x_1,x_2, \ldots,x_{M})$-code. Hence, $$N^{(1)}_{ID}(\boldsymbol{I}_f^{(1)}(t, x_1,x_2, \ldots,x_{M})) \leq N^{(1)}_{ID}(\boldsymbol{I}_f^{(1)}(t, x_1,x_2, \ldots,x_{2^k})).$$ From Theorem~\ref{Theorem_opt_irr} we get $r^f_{ID}(k, t) \geq N^{(1)}_{ID}(\boldsymbol{I}_f^{(1)}(t, x_1,x_2, \ldots,x_{M})).$
         
        \medskip
        \noindent Since $|\mathrm{Im}(f)| \geq 2$ for a non-constant function, by \cite[Corollary 1]{10132545} there exist $x, x' \in \mathbb{F}_2^k$ with $d_H(x,x') = 1$ and $f(x) \neq f(x')$. Using the inequality, $2 \leq d_{ID}(x, x') \leq 2d_H(x,x')$, we get $d_{ID}(x,x') = 2$, that is, there always exist $x, x' \in \mathbb{F}_2^k$ with $d_{ID}(x,x') = 2$ such that $f(x) \neq f(x')$ whenever $|\mathrm{Im}{(f)}| \geq 2.$ Hence, in particular for $M = 2$, we can say $r^f_{ID}(k, t) \geq N^{(1)}_{ID}(2, 2t)$. Consider the following repetition code of length $t$, $\mathcal{C} = \{(0,0, \ldots, 0), (1,1, \ldots, 1)\}$; then $d_{ID}(\mathcal{C}) = 2t$. Hence, $N^{(1)}_{ID}(2, 2t) = t.$
    \end{proof}

    \begin{remark}\label{rmk:singleton-bij}
    When $f$ is bijective, $|\mathrm{Im}(f)| = 2^k$ and $f(x_i)\neq f(x_j)$ for all $x_i \neq x_j$, so the FCIDC codewords form a classical $t$-insdel code with $2^k$ codewords with length $n = k+r$, and minimum insdel distance at least $2t+2$. The insdel Singleton bound (Lemma~\ref{singleton}) then gives $2^k \le 2^{\,(k+r)-(t+1)+1}$, i.e., \ $r \ge t$, which is exactly the lower bound $r^f_{ID}(k,t)\ge t$ of Corollary~\ref{lower_bound_red}. Thus, in the bijective case, this bound coincides with the insdel Singleton bound.
    \end{remark}
    As pointed out in \cite{10132545}, finding the optimal length of irregular-distance codes over a complete set of message vectors is quite difficult. In order to obtain a computationally simpler bound than Theorem~\ref{Theorem_opt_irr}, we define the concept of function distance and function distance matrices.
    
    \begin{definition}[Function Distance] For \(f_1, f_2 \in \mathrm{Im}(f)\), the minimum insdel distance between two information vectors that evaluate to \( f_1 \) and \( f_2 \)  is defined as the insdel distance between the function values $f_1$ and $f_2$, i.e.,
        \[
         d_{ID}^f (f_1, f_2) = \min_{x_1, x_2 \in \mathbb{F}_2^k} d_{ID}(x_1, x_2) \quad \text{ s.t. } f(x_1) = f_1 \text{ and } f(x_2) = f_2.
        \]
    \end{definition}


        \begin{definition}[Function Distance Matrices] 
        \label{def:function-dis-matrices}The function insdel-distance matrices for function $f$ are square matrices of order $E  = |\mathrm{Im}(f)|$ denoted respectively by $\boldsymbol{I}_f^{(1)}(t,f_1,\ldots, f_E)$ and $\boldsymbol{I}_f^{(2)}(t,f_1,\ldots, f_E)$, whose entries are given by:
        {\footnotesize
            \begin{align*}
            &[\boldsymbol{I_f^{(1)}}( t, f_1, \dots, f_E)]_{ij} = 
            \begin{cases} 
                [2(t + 1) - d_{ID}^f (f_i, f_j)]^{+}, & \text{if } i \neq j, \\ 
                0, & \text{otherwise}.
            \end{cases}\\
            &\text{and}\\
            &[\boldsymbol{I_f^{(2)}}( t, f_1, \dots, f_E)]_{ij} = 
            \begin{cases} 
                [2(t + 1 + k)  - d_{ID}^f (f_i, f_j)]^{+}, & \text{if } i \neq j, \\ 
                0, & \text{otherwise}.
            \end{cases}
        \end{align*}
        }       
        \end{definition}

        \begin{theorem}\label{upper_bound_red}
        For an arbitrary function $f : \mathbb{F}_2^k \to \mathrm{Im}(f), \mathrm{Im}(f) = 
        \{f_1,f_2, \ldots ,f_E \}$, where $E = |\mathrm{Im}(f)|$, we have 
        \[
        r_{ID}^f(k,t) \leq N_{ID}^{(2)}\big(\bm{I}_f^{(2)}(t, f_1, \ldots, f_E);k\big).
        \]
        \end{theorem}
        \begin{proof}
        We present a construction that achieves the claimed redundancy. Let $P = \{p_1, p_2, \ldots, p_E\} \subseteq \mathbb{F}_2^{r}$ be a 
        type~2 irregular insdel-distance code for the matrix $I^{(2)}_f(t, f_1, \ldots, f_E)$. Hence, by definition of type~2 irregular insdel-distance code, $r \geq k$ and
        \[
        d_{ID}(p_i, p_j)
        \geq \big[\, 2t + 2 + 2k - d^{f}_{ID}(f_i, f_j) \,\big]^{+} 
        \qquad \text{for all } i \neq j,
        \]
        where no constraint is imposed when $i = j$, as the diagonal entries of $I^{(2)}_f$ are zero. Assign to each $x \in \mathbb{F}_2^k$ with $f(x) = f_i$ the common redundancy vector $p_i$, and define $\psi(x) = (x, p_i)$.
        
        \noindent
        Let $x_i, x_j \in \mathbb{F}_2^k$ satisfy $f(x_i) = f_i$, $f(x_j) = f_j$, with $f_i \neq f_j$. Since $r \geq k$, we have $\min\{k, r\} = k$, and Lemma~\ref{insdel_inequality} gives
        \[
            d_{ID}(\psi(x_i), \psi(x_j)) 
            \geq d_{ID}(x_i, x_j) + d_{ID}(p_i, p_j) - 2k.
        \]
        Applying the distance condition on $\{p_i, p_j\}$, we obtain
        \[
        d_{ID}(\psi(x_i), \psi(x_j)) 
        \geq d_{ID}(x_i, x_j) + 
       \big(2t + 2 + 2k - d^{f}_{ID}(f_i, f_j)\big) - 2k
        = d_{ID}(x_i, x_j) + 2t + 2 - d^{f}_{ID}(f_i, f_j).
        \]
        By definition of the function distance, $d^{f}_{ID}(f_i, f_j) \leq d_{ID}(x_i, x_j)$, since $(x_i, x_j)$ constitutes one admissible pair in this minimization. Consequently,
         \[
            d_{ID}(\psi(x_i), \psi(x_j)) 
            \geq d_{ID}(x_i, x_j) + 2t + 2 -  d^{f}_{ID}(f_i, f_j) 
            \geq 2t + 2.
        \]
        Thus, $\psi$ defines a function-correcting insdel code for $f$ with redundancy $r$. Minimizing over all such type~2 codes yields
        \[
            r^{f}_{ID}(k, t) 
             \leq N^{(2)}_{ID}\!\big(I^{(2)}_f(t, f_1, \ldots, f_E); 
             k\big). 
        \]
        \end{proof}

        It is not always easy to derive a generic formula for the function distance $d_{ID}^f(f_i, f_j)$. Consequently, an upper bound on the optimal redundancy can be estimated using a lower constraint on $d_{ID}^f(f_i, f_j)$ in the absence of an explicit equation.
        
        \begin{lemma}\label{lower_bound_function_distance}
            Let $f: \mathbb{F}_2^k \rightarrow \mathrm{Im}(f)$ be a function with expressiveness $E = |\mathrm{Im}(f)|$ and let $\{a_{ij}\}_{1 \leq i, j \leq E}$ be an $E \times E$ matrix whose entries satisfy
            \[
            a_{ij} = a_{ji} \quad \text{and} \quad 
            a_{ij} \leq d^{f}_{ID}(f_i, f_j) 
            \qquad \text{for all } 1 \leq i, j \leq E.
            \]
            Then,
            $$N_{ID}^{(2)}(\bm{I}_f^{(2)}(t, f_1, f_2, \ldots, f_E);k) \leq N_{ID}^{(2)}(I;k)$$
            where $I$ is a symmetric square matrix of order $E$ whose entries are given by 
            \begin{align*}
             [{\mathbf{I}}]_{ij} =
            \begin{cases} 
                [2t + 2 + 2k - a_{ij}]^{+}, & \text{if } i \neq j, \\ 
                0, & \text{otherwise}.
            \end{cases}
            \end{align*}
        \end{lemma}
        \begin{proof}
            Let $\mathcal{C} = \{ c_1, c_2, \ldots, c_E\}$ be a length $N_{ID}^{(2)}(I;k)$, irregular insdel-distance code of type $2$ corresponding to matrix $I$. Then, for all $i \neq j$
            \begin{align*}
                d_{ID}(c_i, c_j) &\geq [2t + 2+ 2k - a_{ij}]^+\\
                                     & \geq [2t + 2+ 2k - d_{ID}^f(f_i, f_j)]^+\\
                                     & = [\boldsymbol{I}_f^{(2)}( t, f_1, \dots, f_E)]_{ij}
            \end{align*}
            Hence, $\mathcal{C}$ is an irregular insdel-distance code of type $2$ for function distance matrix ${\bm{I}_f^{(2)}}(t,f_1,\ldots, f_E)$.
            Therefore, $N_{ID}^{(2)}(\bm{I}_f^{(2)}(t, f_1, f_2, \ldots, f_E);k) \leq N_{ID}^{(2)}(I;k).$
        \end{proof}
        
        Combining Lemma~\ref{lower_bound_function_distance} and Theorem~\ref{upper_bound_red} yields the following corollary.
        
        \begin{corollary}\label{optional_red_and_function_dis_bound}
            Let $f: \mathbb{F}_2^k \rightarrow \mathrm{Im}(f)$ be a function with expressiveness $E = |\mathrm{Im}(f)|$ and let $\{a_{ij}\}_{1 \leq i, j \leq E}$ be an $E \times E$ matrix whose entries satisfy $a_{ij} = a_{ji}$ and $a_{ij} \leq d^{f}_{ID}(f_i, f_j)$ for all $1 \leq i, j \leq E$. Then,
            $$r_{ID}^f(k,t) \leq N_{ID}^{(2)}(\bm{I};k)$$
            where $\bm{I}$ is a symmetric square matrix of order $E$ whose entries are given by 
            \begin{align*}
             [\bm{I}]_{ij} =
            \begin{cases} 
                [2t + 2 + 2k - a_{ij}]^{+}, & \text{if } i \neq j, \\ 
                0, & \text{otherwise}.
            \end{cases}
            \end{align*}
        \end{corollary}


    The next lemma is derived from the proof of \cite[Lemma 1]{hayashi2020list} by assuming the existence of a common binary supersequence $v$ of length $N$ for all codewords in the binary insdel code, and setting $t_D = 0$ (number of deletions) and $t_I = N - n$ (number of insertions), where $n$ is the length of the codewords.

    \begin{lemma}\label{lem:superseq-plotkin}
        Let $C=\{c_1,\dots,c_M\}\subseteq\mathbb{F}_2^{n}$ be any set of $M$ binary words of common length $n$. Suppose there exists $v\in\mathbb{F}_2^{N}$, $N\ge n$, that contains every word of $C$ as a subsequence. Then
        \[
        \sum_{1\le i<j\le M} d_{ID}(c_i,c_j)\;\le\;\frac{M^{2}(N-n)\,n}{N}.
        \]
    \end{lemma}

    \begin{remark}\label{rem:superseq-existence}
    A common supersequence $v\in\mathbb{F}_2^{N}$ of $C$ exists if and only if $N\ge \mathrm{SCS}(C)$, where $\mathrm{SCS}(C)$ denotes the length of a shortest common supersequence of $C$. Also, $n\le \mathrm{SCS}(C)\le Mn$, where the upper bound can be obtained by considering the concatenation of all the codewords of $C$ ($c_1c_2\cdots c_M$). Therefore, the bound in Lemma~\ref{lem:superseq-plotkin} is vacuous for $N<\mathrm{SCS}(C)$ and is applied only for $N$ in the admissible range.
    \end{remark}

    Using the result from the previous lemma, we will now establish a lower bound on $N_{ID}^{(1)}(I)$ for a given distance matrix $I$. This bound can be viewed as a generalization of the Plotkin-type bound for insdel codes with irregular distance requirements.

    \begin{lemma}\label{plotkin-like-bound}
    Let $\mathbf{I}\in\mathbb{N}_0^{M\times M}$ and put $S=\sum_{i<j}[\mathbf{I}]_{ij}$. For every $N\in\mathbb{N}$ with $N\ge \max\{\mathrm{SCS}(\mathcal{P}),\,4S/M^{2}\}$, where $\mathcal{P}$ is a shortest type 1 irregular insdel-distance code for $\mathbf{I}$,
    \[
        N_{ID}^{(1)}(\mathbf{I})\;\ge\;\frac{N-\sqrt{\,N^{2}-\dfrac{4SN}{M^{2}}\,}}{2}.
    \]
    \end{lemma}

    \begin{proof}
    Let $\mathcal{P}=\{p_1,\dots,p_M\}\subseteq\mathbb{F}_2^{r}$ be a type 1 irregular insdel-distance code corresponding to $\mathbf{I}$ such that its length satisfies $r=N_{ID}^{(1)}(\mathbf{I})$. The concatenation $p_1p_2\cdots p_M$ is a common supersequence of $\mathcal{P}$, so by Remark~\ref{rem:superseq-existence} a common supersequence $v\in\mathbb{F}_2^{N}$ exists for every $N\ge \mathrm{SCS}(\mathcal{P})$.
    Fix any such $N$ that additionally satisfies $N\ge 4S/M^{2}$.

    Since $\mathcal{P}$ is a set of $M$ words of common length $r$, Lemma~\ref{lem:superseq-plotkin} gives
    \[
  \sum_{i<j} d_{ID}(p_i,p_j)\;\le\;\frac{M^{2}(N-r)\,r}{N}.
    \]
    By the type 1 distance constraint $d_{ID}(p_i,p_j)\ge[\mathbf{I}]_{ij}$, we have $S=\sum_{i<j}[\mathbf{I}]_{ij}\le \sum_{i<j} d_{ID}(p_i,p_j)$. Combining the above two inequalities, we get
    \[
    S\le \frac{M^{2}(N-r)\,r}{N}
    \;\Longleftrightarrow\;
    r^{2}-Nr+\frac{SN}{M^{2}}\le 0,
    \]
    which gives
    \[
    r\;\ge\;\frac{N-\sqrt{\,N^{2}-\frac{4SN}{M^{2}}\,}}{2}.
    \]
    As $r=N_{ID}^{(1)}(\mathbf{I})$, the claim follows.
    \end{proof}

    The bound of Lemma~\ref{plotkin-like-bound} is conditional, since it requires a common supersequence of the optimal code to exist. The following lemma removes this dependence entirely by using $d_{H}(p_i,p_j)\ge\tfrac12\,d_{ID}(p_i,p_j)$ together with the Hamming Plotkin-type bound of Lemma~\ref{plotkin_FCC}, yielding the same estimate unconditionally. We therefore adopt Lemma~\ref{plotk} as our primary Plotkin-type bound.

    \begin{lemma}
    \label{plotk}
    For any insdel distance-constraint matrix $\mathbf{I} \in \mathbb{N}_0^{M \times M}$, we have
    \[
    N^{(1)}_{ID}(\bm{I}) \;\geq\; 
    \begin{cases}
        \dfrac{2}{M^2} \displaystyle\sum_{i < j} [\bm{I}]_{ij}, 
        & \text{if } M \text{ is even}, \\[1.2ex]
        \dfrac{2}{M^2 - 1} \displaystyle\sum_{i < j} [\bm{I}]_{ij}, 
        & \text{if } M \text{ is odd}.
    \end{cases}
    \]
    \end{lemma}

    \begin{proof}
        Let $P = \{p_1, p_2, \ldots, p_M\} \subseteq \mathbb{F}_2^{r}$ be a type~1 irregular insdel-distance code for $\bm{I}$ of length $r = N^{(1)}_{ID}(\bm{I})$, so that $d_{ID}(p_i, p_j) \geq [\bm{I}]_{ij}$ for all $i \neq j$.

        \smallskip
        \noindent
        By Lemma~\ref{lem:rel_insdel_hamming}, $d_H(p_i, p_j) \geq \tfrac{1}{2} d_{ID}(p_i, p_j) \geq \tfrac{1}{2} [\bm{I}]_{ij}$, hence
        \[
         d_H(p_i, p_j) \;\geq\; \big\lceil \tfrac{1}{2}[I]_{ij} \big\rceil \qquad \text{for all } i \neq j.
        \]
        Define matrix $D \in \mathbb{N}_0^{M \times M}$ by
        \[
        [D]_{ij} \;=\; 
        \begin{cases}
        \big\lceil \tfrac{1}{2}[I]_{ij} \big\rceil, & i \neq j, \\
        0, & i = j.
        \end{cases}
        \]
        Then by Definition~\ref{def:irr-dis-code}, $P$ is an irregular-distance code for matrix $D$. Hence, by Lemma~\ref{plotkin_FCC},
        \[
        r \;\geq\; N(D) \;\geq\; 
        \begin{cases}
        \dfrac{4}{M^2} \displaystyle\sum_{i<j} [D]_{ij}, 
        & \text{if } M \text{ is even}, \\[1.2ex]
        \dfrac{4}{M^2 - 1} \displaystyle\sum_{i<j} [D]_{ij}, 
        & \text{if } M \text{ is odd}.
    \end{cases}
        \]
        Since
        \[
        \sum_{i<j} [D]_{ij} 
        = \sum_{i<j} \big\lceil \tfrac{1}{2}[\bm{I}]_{ij} \big\rceil 
         \;\geq\; \tfrac{1}{2} \sum_{i<j} [\bm{I}]_{ij},
        \]
        the bound on $r$ becomes
        \[
        r \;\geq\; 
        \begin{cases}
        \dfrac{4}{M^2} \cdot \tfrac{1}{2} \displaystyle\sum_{i<j} [\bm{I}]_{ij} 
        = \dfrac{2}{M^2} \displaystyle\sum_{i<j} [\bm{I}]_{ij}, 
        & M \text{ even}, \\[1.4ex]
        \dfrac{4}{M^2 - 1} \cdot \tfrac{1}{2} \displaystyle\sum_{i<j} [\bm{I}]_{ij} 
        = \dfrac{2}{M^2 - 1} \displaystyle\sum_{i<j} [\bm{I}]_{ij}, 
        & M \text{ odd}.
    \end{cases}
        \]
    Since this holds for every type~1 irregular insdel-distance code for matrix $I$, the bound holds for $N^{(1)}_{ID}(I)$.
    \end{proof}

    \begin{corollary}
    \label{cor:plotkin-insdel}
    Let $M$ be a positive integer and $d$ an even positive integer, and let $\mathbf{I}$ be the matrix with $[\mathbf{I}]_{ij} = d$ for all $i \neq j$. Then Lemma~\ref{plotk} gives the Plotkin bound for classical insdel codes,
    \[
    N^{(1)}_{ID}(M, d) \;\geq\;
    \begin{cases}
    \dfrac{(M-1)\,d}{M}, & M \text{ even},\\[1.4ex]
    \dfrac{M\,d}{M+1}, & M \text{ odd}.
    \end{cases}
    \]
    In particular, if $f : \mathbb{F}_2^k \to \mathrm{Im}(f)$ is bijective, then $f(x_i) \neq f(x_j)$ for all $x_i \neq x_j$, so any $t$-insdel error-correcting FCIDC for $f$ is a classical $t$-insdel error-correcting code $C \subseteq \mathbb{F}_2^{n}$, $n = k + r$, with $|C| = 2^k$ and $d_{ID}(C) \geq 2t + 2$. Taking $M = 2^k$ (even) and $d = 2t+2$ above yields
    \[
    n \;\geq\; \frac{(2^k - 1)(2t + 2)}{2^k},
    \]
    the classical Plotkin bound for insdel codes. Thus, the Plotkin-like bound for irregular insdel-distance codes specializes to the insdel Plotkin bound whenever $f$ is bijective.
    \end{corollary}

    \begin{remark}
    The bound of Corollary~\ref{cor:plotkin-insdel} coincides with the Plotkin-type bound for the insdel metric of Hayashi and Yasunaga~\cite[Thm.~4]{hayashi2020list}. They likewise note that this bound follows from the Hamming Plotkin bound via $d_H \ge \tfrac12 d_{ID}$, which is exactly the route taken in Lemma~\ref{plotk}. The even/odd refinement here is inherited from the irregular-distance Plotkin bound of~\cite{10132545}.
    \end{remark}

    The next lemma is a variant of the Gilbert-Varshamov bound for irregular insdel-distance codes.
    
    \begin{lemma}\label{Lemma Gv}
        For any insdel distance-constraint matrix \( \mathbf{I} \in \mathbb{N}^{M\times M}_0 \) and any permutation 
        \( \pi : \{1, 2, \ldots, M\} \rightarrow \{1, 2, \ldots, M\} \), we have
        \begin{align*}
         &N^{(1)}_{ID}(\boldsymbol{I}) \leq \min_{r \in \mathbb{N}} \left\{ r \ \middle|\ 2^r > \max_{j \in \{1, 2, \ldots, M\}} \sum_{i=1}^{j-1} B^{\max}_{ID}\big(r,[\boldsymbol{I}]_{\pi(i)\pi(j)} - 2\big) \right\},\\       &\text{ and } \\&N^{(2)}_{ID}(\boldsymbol{I};K) \leq \min_{r \geq K} \left\{ r \ \middle|\ 2^r > \max_{j \in \{1, 2, \ldots, M\}} \sum_{i=1}^{j-1} B^{\max}_{ID}\big(r,[\boldsymbol{I}]_{\pi(i)\pi(j)} - 2\big) \right\}.  
        \end{align*}
        \[
            B^{\max}_{ID}(r,t)
            \triangleq
            \max_{x \in \mathbb{F}_2^r}
            \left| B_{ID}(x,t) \right|,
            \qquad
            B_{ID}(x,t) = \{\, y \in \mathbb{F}_2^r \mid d_{ID}(x,y) \le t \,\}.
        \]
    \end{lemma}
    \begin{proof}
        We present an iterative way of selecting valid codewords for constructing an irregular insdel-distance code of type~2 (type~1) for any insdel distance-constraint matrix $\mathbf{I} \in \mathbb{N}_0^{M \times M}$.
        For simplicity, first choose $\pi$ as the identity permutation.
        Select an arbitrary vector $p_1 \in \mathbb{F}_2^r$ as the first codeword, where $r \ge K$ (or $r \in \mathbb{N}$ for type 1).
        For the second codeword $p_2$, the distance requirement 
        \[
            d_{ID}(p_1,p_2) \ge [\boldsymbol{I}]_{12}
        \]
        must be satisfied. Such a $p_2$ exists provided
        \[
            2^r > \left| B_{ID}\big(p_1,[\boldsymbol{I}]_{12}-2\big) \right|.
        \]
        Since the size of an insdel ball depends on its center, we upper bound it by the worst-case ball size $B^{\max}_{ID}(r,[\boldsymbol{I}]_{12}-2)$.

        Next, choose the third codeword $p_3$ such that
        \[
            d_{ID}(p_1,p_3) \ge [\boldsymbol{I}]_{13}
            \quad \text{and} \quad
            d_{ID}(p_2,p_3) \ge [\boldsymbol{I}]_{23}.
        \]
        A sufficient condition for the existence of such a $p_3$ is
        \[
            2^r >
            \left| B_{ID}\big(p_1,[\boldsymbol{I}]_{13}-2\big) \right| + \left| B_{ID}\big(p_2,[\boldsymbol{I}]_{23}-2\big) \right|,
        \]
        which is guaranteed if
        \[
            2^r > B^{\max}_{ID}\big(r,[\bm{I}]_{13}-2\big) + B^{\max}_{ID}\big(r,[\bm{I}]_{23}-2\big).
        \]
        Proceeding inductively, at the $j$-th step we must choose $p_j$ such that
        \[
        d_{ID}(p_i,p_j) \ge [\boldsymbol{I}]_{ij}
        \quad \text{for all } i<j.
        \]
        A sufficient condition for the existence of $p_j$ is
        \[
            2^r >
            \sum_{i=1}^{j-1}
            B^{\max}_{ID}\big(r,[\boldsymbol{I}]_{ij}-2\big).
        \]

        Since the codewords can be selected in any order, the same argument holds for any permutation $\pi$, completing the proof.
    \end{proof}
    
    \begin{corollary}
    \label{cor:bijective-GV}
    Let $M$ be a positive integer and $d$ an even positive integer, and let $\mathbf{I}$ be the matrix with $[\mathbf{I}]_{ij}=d$ for all $i\neq j$. Then the Gilbert-Varshamov-type bound of Lemma~\ref{Lemma Gv} gives, for the classical insdel code parameters,
    \[
    N^{(1)}_{ID}(M,d)\;\le\;
    \min_{N\in\mathbb{N}}\Bigl\{\,N \;\Big|\; 2^{N}>(M-1)\,B^{\max}_{ID}(N,d-2)\,\Bigr\}.
    \]
    In particular, if $f:\mathbb{F}_2^k\to\mathrm{Im}(f)$ is bijective, then $f(x_i)\neq f(x_j)$ for all $x_i\neq x_j$, so any $t$-insdel-correcting FCIDC for $f$ is a classical $t$-insdel-correcting code $C\subseteq\mathbb{F}_2^{n}$, $n=k+r$, with $|C|=2^k$ and $d_{ID}(C)\geq 2t+2$. Taking $M=2^k$ and $d=2t+2$ above, the redundancy satisfies
    \[
    r \;=\; N^{(1)}_{ID}(2^k,2t+2)-k \;\leq\; 2t\log_2 k + O_t(1).
    \]
    \end{corollary}

    The asymptotic redundancy $r\le 2t\log_2 k + O_t(1)$ obtained in the Corollary~\ref{cor:bijective-GV} matches the classical Gilbert-Varshamov redundancy bound for classical $t$-insdel-correcting codes \cite{levenshtein1965двоичные}.
    \begin{theorem}\label{N_M_d_bound}
        For a positive integer $M$, an even positive integer $d$, and $K \geq 2$, we have
        \[
            N_{ID}^{(2)}(M,d;K) \le \Bigg\lceil \frac{\ln M + (d-2)\ln\!\Big(\frac{2eK}{d-2}\Big)}{\ln 2}\Bigg\rceil.
        \]
    \end{theorem}

    \begin{proof}
        By Lemma~\ref{Lemma Gv}, the minimum length of a type~2 irregular insdel-distance code with $M$ codewords and distance requirement $d$ satisfies
        \begin{align*}
            N^{(2)}_{ID}(M,d;K) &\le \min_{r \ge K} \left\{r \,\middle|\, 2^r > \max_{j \in \{1,\ldots,M\}} \sum_{i=1}^{j-1} B^{\max}_{ID}(r,d-2)\right\} \\
            &\le \min_{r \ge K}\left\{r \,\middle|\, 2^r > M \, B^{\max}_{ID}(r,d-2)\right\},
        \end{align*}
        where
        \[
            B^{\max}_{ID}(r,t)
            \triangleq
            \max_{x \in \mathbb{F}_2^r}
            \left| B_{ID}(x,t) \right|.
        \]

        Using the upper bound on the insdel ball size from Lemma~\ref{insdel_ball}, we have
        \[
        B^{\max}_{ID}(r,d-2)
        \le
        \frac{
        e^{d-2}\,r^{\frac{d-2}{2}}\,
        \left(r+\frac{d-2}{2}\right)^{\frac{d-2}{2}}}{\left(\frac{d-2}{2}\right)^{d-2}}.
        \]

        Let $d' = d-2$. A sufficient condition for
        \(M  \cdot\, B^{\max}_{ID}(r,d') < 2^r\)
        is
        \[
            \frac{
            e^{d'}\,r^{\frac{d'}{2}}\,
            \left(r+\frac{d'}{2}\right)^{\frac{d'}{2}}}{\left(\frac{d'}{2}\right)^{d'}}\, M < 2^r.
        \]

        Taking natural logarithms on both sides yields
        \[
            d' + \frac{d'}{2}\ln r
            + \frac{d'}{2}\ln\!\left(r+\frac{d'}{2}\right)
            - d'\ln\!\left(\frac{d'}{2}\right)
            + \ln M
            < r \ln 2.
        \]

        Using
        \(
            -d'\ln\!\left(\frac{d'}{2}\right)
            = -d'\ln d' + d'\ln 2
        \)
        and the inequality
        \(
        \ln\!\left(r+\frac{d'}{2}\right) \ge \ln r,
        \)
        we obtain
        \[
            \ln M + d' - d'\ln d' + d'\ln 2 < r\ln 2 - d'\ln r.
        \]

        Since $r \ge K$, we have $\ln r \ge \ln K$, and hence
        \[
        r\ln 2 > \ln M + d' + d'\ln\!\Big(\frac{2K}{d'}\Big).
        \]
        Dividing both sides by $\ln 2$ gives
        \[
            r > \frac{ \ln M + d'\ln\!\Big(\frac{2eK}{d'}\Big) }{\ln 2}.
        \]
        Therefore,
        \[
        N_{ID}^{(2)}(M,d;K)
        \le
        \Bigg\lceil
        \frac{
        \ln M + (d-2)\ln\!\Big(\frac{2eK}{d-2}\Big)}{\ln 2}\Bigg\rceil,
        \]
        which completes the proof.
    \end{proof}

    \begin{theorem}
    \label{N1_M_d_bound}
    For a positive integer $M$ and an even positive integer $d$, we have
    \[
    N^{(1)}_{ID}(M,d) \le \max\left\{
    \frac{2\ln M + (d-2)\ln 8}{\ln 2},
    \frac{4(d-2)}{\ln 2} \ln\!\left(\frac{4(d-2)}{\ln 2}\right)
    \right\}.
    \]
    \end{theorem}

    \begin{proof}
    We follow the same initial steps as in the proof of Theorem~\ref{N_M_d_bound}. Let $d' = d-2$. A sufficient condition for the existence of a type~1 code of length $r$ is
    \[
    2^r > M \cdot B^{\max}_{ID}(r,d'),
    \]
where $B^{\max}_{ID}(r,t)$ denotes the maximum size of an insdel ball of radius $t$ in $\mathbb{F}_2^r$.
\noindent By Lemma~\ref{insdel_ball}, we have the bound
\[
B^{\max}_{ID}(r,d') \le \left(\frac{2e\sqrt{r(r+d'/2)}}{d'}\right)^{d'}.
\]
Since any code of length $r$ and minimum distance $d$ must satisfy $d \le 2r$, we have $d' = d-2 \le 2r-2 < 2r$. Thus $r+d'/2 \le 2r$, and consequently
\[
B^{\max}_{ID}(r,d') \le \left(\frac{2e\sqrt{r \cdot 2r}}{d'}\right)^{d'}
= \left(\frac{2\sqrt{2}\,e\,r}{d'}\right)^{d'}
\le \left(\frac{8r}{d'}\right)^{d'}.
\]
\noindent It therefore suffices to find $r$ such that
\[
    2^r > M \left(\frac{8r}{d'}\right)^{d'},
\]
or equivalently,
\[
r \ln 2 > \ln M + d' \ln(8r) - d'\ln d'.
\]
Rearranging terms gives
\[
r > a + b \ln r,
\]
where
\[
a \triangleq \frac{\ln M + d'\ln 8}{\ln 2}, \qquad
b \triangleq \frac{d'}{\ln 2}.
\]
\noindent In order to solve the inequality $r > a + b\ln r$, we use the observation that if $r \ge 2a$ and $r \ge 4b\ln(4b)$, then the right-hand side is at most $r$.
\begin{itemize}
\item If $r \ge 2a$, then $a \le r/2$.
\item If $r \ge 4b\ln(4b)$, then $b\ln r \le r/2$.
  Equivalently, $r/\ln r \ge 2b$. The function $g(r) := r/\ln r$ is
  increasing for $r \ge e$, and $4b\ln(4b) \ge e$ in our range
  (indeed $4b \ge e$ since $b = (d-2)/\ln 2 \ge 2/\ln 2$), so it suffices
  to verify the bound at the endpoint $r_0 = 4b\ln(4b)$:
  \[
    g(r_0) = \frac{4b\ln(4b)}{\ln(4b) + \ln\ln(4b)}
           \;\ge\; \frac{4b\ln(4b)}{2\ln(4b)} = 2b,
  \]
  where the middle inequality uses $\ln\ln(4b) \le \ln(4b)$. Since $g$ is non-decreasing on $[r_0,\infty)$, we get $g(r) \ge g(r_0) \ge 2b$ for all $r \ge r_0$, i.e. $b\ln r \le r/2$.
\end{itemize}
Thus, any $r$ satisfying
\begin{equation}
\label{eq: bound-on-r}
    r \ge \max\{2a,\; 4b\ln(4b)\}
\end{equation}
also satisfies $r \ge a + b\ln r$.
Substituting the definitions of $a$ and $b$ into \eqref{eq: bound-on-r} yields
\[
r \ge \max\left\{
\frac{2(\ln M + d'\ln 8)}{\ln 2},\;
\frac{4d'}{\ln 2}\ln\!\left(\frac{4d'}{\ln 2}\right)
\right\}.
\]
Since $N^{(1)}_{ID}(M,d)$ is the smallest such $r$, we obtain
\[
N^{(1)}_{ID}(M,d) \le
\max\left\{
\frac{2(\ln M + d'\ln 8)}{\ln 2},\;
\frac{4d'}{\ln 2}\ln\!\left(\frac{4d'}{\ln 2}\right)
\right\} .
\]

\end{proof}

    To bound $N_{ID}^{(2)}(\mathbf{I};K)$, we first relate it to $N_{ID}^{(2)}(M,I;K)$. The bound on the latter, provided by Theorem~\ref{N_M_d_bound}, then directly implies a bound on the former.
    
    \begin{corollary}\label{bound_Imax}
    Let ${I} \in \mathbb{N}_0^{M\times M}$, and let $I_{max}$ denote its largest entry. If $I_{max} \leq I$, then
    \[
        N_{ID}^{(2)}({I};K) \leq N_{ID}^{(2)}(M,I;K).
    \]
    \end{corollary}


\section{VT-Syndrome Function}
    The VT-syndrome function captures the weighted sum of bit positions for a given binary sequence and is used to define a VT Code that is used to correct single insdel errors.
    
    \begin{definition}[VT-syndrome function]
        A VT-syndrome function is defined as $f(u_i) = VT(u_i) = \sum_{j = 1}^{k} ju_{ij} \pmod{k+1}$, $\forall$ $u_i = (u_{i1}, \ldots, u_{ik}) \in \mathbb{F}_2^k$ \text{ and }  $\forall$ $i \in [2^k]$ \text{ such that } $E = |\mathrm{Im}(f)| = k+1.$
    \end{definition}

    Using Theorem~\ref{upper_bound_red} and Corollary~\ref{lower_bound_red}, we establish an upper bound and lower bound on $r^{VT}_{ID}(k,t)$ in the next lemma.
    
    \begin{lemma}\label{VT_red_bound}
        Let \( f: \mathbb{F}_2^k \rightarrow \{0, 1, \ldots,k\}\) be the VT-syndrome function defined by \( f_i = i - 1 \) for all \( 1 \leq i \leq k + 1 \), and let $x_1 = (0, \ldots, 0)$, \( x_i = (0^{i-2} 1 0^{k - i + 1}) \) denote \( k + 1 \) representative binary vectors of length $k$, where \( 2 \leq i \leq k + 1 \). Consider the function distance matrix \( \bm{I}_f^{(2)}(t, f_1, \ldots, f_{k+1}) \) corresponding to \( f \), and the distance matrix \( \bm{I}_{f}^{(1)}(t, x_1, \ldots, x_{k+1}) \) corresponding to the set of representative binary vectors.  
        Then,
        
        \[
            N_{ID}^{(1)}(\bm{I}_{f}^{(1)}(t, x_1, \ldots, x_{k+1})) 
            \;\leq\;
            r^{\text{VT}}_{ID}(k, t)
            \;\leq\;
            N_{ID}^{(2)}(\bm{I}_f^{(2)}(t, f_1, \ldots, f_{k+1}); k),
        \]
        
        where
        
        \begin{align*}
            &[\bm{I}_f^{(2)}((t, f_1, \ldots, f_{k+1}); k)]_{ij} =
            \begin{cases}
                0,      & i = j, \\[4pt]
                2(t + k), & i \neq j,
            \end{cases}\\
            &\text{ and }\\
            &[\bm{I}_{f}^{(1)}(t, x_1, \ldots, x_{k+1})]_{ij} =
            \begin{cases}
                0,  & i = j, \\[4pt]
                2t, & i \neq j.
            \end{cases}
        \end{align*}
        
    \end{lemma}
        
    \begin{proof}
        Consider the set of \( k + 1 \) binary vectors \( \{x_i\}_{i=1}^{k+1} \) defined by
        \[
            x_i = (0, 0, \ldots,\underbrace{1}_{(i-1)\text{th position}},\ldots, 0), \quad 1 \leq i \leq k + 1.
        \]
        The function values corresponding to these vectors are given by
        \[
            f_i = f(x_i) = i - 1, \quad \forall\, 1 \leq i \leq k + 1.
        \]
        For any distinct pair \( (x_i, x_j) \), the insdel-distance satisfies
        \[
            d_{ID}(x_i, x_j) = 2(k - (k - 1)) = 2.
        \]
        Hence, the corresponding function distance satisfies
        \[
            2 \leq d_{ID}^{f}(f_i, f_j) \leq d_{ID}(x_i, x_j) = 2, 
            \quad \forall\, 1 \leq i \neq j \leq k + 1.
        \]
        Therefore, the entries of the function insdel matrix and insdel-distance matrices associated with the set \( \{x_i\}_{i=1}^{k+1} \) are given by
        \[
            [\bm{I}_f^{(2)}((t, f_1, \ldots, f_{k+1}); k)]_{ij} =
            \begin{cases}
                0,       & i = j, \\[4pt]
                2(t + k), & i \neq j,
            \end{cases}
        \]
        and
        \[
            [\bm{I}_{f}^{(1)}(t, x_1, \ldots, x_{k+1})]_{ij} =
            \begin{cases}
                0,  & i = j, \\[4pt]
                2t, & i \neq j.
            \end{cases}
        \]
        This establishes the desired bounds on \( r^{\text{VT}}_{ID}(k, t) \), completing the proof.
    \end{proof}
    
    \begin{remark}\label{VT_N_M_d}
        Since the non-diagonal entries of both the distance matrices $I_{f}^{(1)}(t, x_1, \ldots, x_{k+1})$ and $I_{f}^{(2)}(t, f_1, \ldots, f_{k+1})$ are equal to the same value $2t$ and $2(t+k)$ respectively, the bound on the optimal redundancy in Lemma~\ref{VT_red_bound} can be expressed as:
        \[
            N_{ID}^{(1)}(k+1, 2t) 
            \;\leq\;
            r^{\text{VT}}_{ID}(k, t)
            \;\leq\;
            N_{ID}^{(2)}(k+1, 2(t+k); k)
        \]
    \end{remark}

    \noindent
    The next corollary gives an explicit bound on the optimal redundancy of FCIDCs for the VT syndrome function. The upper bound can be derived by using the result of the preceding remark together with Theorem~\ref{N_M_d_bound}, whereas the lower bound follows from Remark~\ref{VT_N_M_d} and Lemma~\ref{plotk}.
    
    \begin{corollary}
        Let $k \geq  2$ and $t \in \mathbb{N}$. Then the optimal redundancy of FCIDCs for the VT syndrome function defined over $\mathbb{F}_2^k$ is bounded as follows:
        \[
            \frac{2tk}{k+1} \leq r^{\text{VT}}_{ID}(k, t) \leq \Bigg \lceil \frac{\ln (k+1) + 2(t+k-1)\ln \Big(\frac{ek}{t+ k - 1}\Big)}{\ln 2} \Bigg \rceil.
        \]
    \end{corollary}
    The following example illustrates Lemma~\ref{VT_red_bound} for particular values of $k$ and $t$.

    \begin{example}
        Let \( k = 2 \) and \( t = 1 \). Then, from Lemma~\ref{VT_red_bound}, the function distance matrix corresponding to the VT-syndrome function and the distance matrix corresponding to the binary vectors $x_1 = (00)$, $x_2 = (10)$, \text{ and } $x_3 = (01)$ are given respectively by
        \[
            \bm{I}_f^{(2)}(1, f_1, f_2, f_3) =
            \begin{pmatrix}
                0 & 6 & 6 \\[4pt]
                6 & 0 & 6 \\[4pt]
                6 & 6 & 0
            \end{pmatrix},
        \]
        and
        \[
            \bm{I}_{\textit{f}}^{(1)}(1, x_1, x_2, x_3) =
            \begin{pmatrix}
                0 & 2 & 2 \\[4pt]
                2 & 0 & 2 \\[4pt]
                2 & 2 & 0
            \end{pmatrix}.
        \]
        Consider the code $\mathcal{C} = \{000000, 000111, 110100\}$. It can be verified that the insdel distance between each pair of codewords in $\mathcal{C}$ is $6$. Hence,
        $$ r^{\text{VT}}_{ID}(2, 1)  \leq N_{ID}^{(2)}(\bm{I}_f^{(2)}(1, f_1, f_2, f_{3}); 2) \leq 6. $$
        \end{example}
\section{Number-Of-Runs Function}
\label{sec:runs}
    
    A run in a binary sequence is a maximal block of consecutive equal bits. One can describe a binary string $x$ in terms of runs, and it is called a run-length sequence, $R(x) = (r_1, r_2, \ldots, r_m)$.

    \begin{definition}
    \label{def:number-of-runs}
        The number-of-runs function is defined as $f(x) = r(x)$, where $r(x) = |R(x)|$, $x \in \mathbb{F}_2^k$, and $k \in \mathbb{N}$.
    \end{definition}

    \begin{example}
        Consider the vector $u = 0100101$, then the run-length sequence of $u$ is $(1,1,2,1,1,1)$ and hence the total number-of-runs equals the length of the run-length vector, i.e., $r(u) = 6$.
    \end{example}

    The expressiveness of this function is given by $E = |\mathrm{Im}(r(\cdot))| = k$. Using Corollary~\ref{lower_bound_red} and Theorem~\ref{upper_bound_red}, we establish bounds on the optimal redundancy of FCIDCs for the function $r(\cdot)$, as stated in the following lemma.

    \begin{lemma}\label{runs_bound}
        Let $f(x) = r(x)$ be the number-of-runs function on $x \in \mathbb{F}_2^k$. Consider the set of $k$ binary vectors defined as $x_i = 0^{k-i+1}(10)^{(i-1)/2}$ if $i$ is odd and $x_i = 0^{k-i+1}(10)^{(i-2)/2}1$ if $i$ is even. Then the optimal redundancy of FCIDCs corresponding to $f$ is bounded as follows:
        
        \[
             N^{(1)}_{ID}(\bm{I}_f^{(1)}(t, x_1,x_2,\dots,x_k)) \leq r_{ID}^f(k,t) \leq N^{(2)}_{ID}(\bm{I}_f^{(2)}(t, f_1,f_2,\dots,f_k);k).
        \]           
                    
            where $\bm{I}_f^{(1)}(t, x_1,x_2,\dots,x_k)$ and $\bm{I}_f^{(2)}(t, f_1,f_2,\dots,f_k)$ are an order $k$ insdel distance matrix and function distance matrix, respectively, and their entries are as follows:
           
            \begin{align*}
            &[\bm{I}_f^{(2)}(t, f_1, \ldots, f_{k})]_{ij} =
            \begin{cases}
                0,      & i = j, \\[4pt]
                2(t + k) + 1 - |i-j|, & i \neq j \text{ and } |i-j| \text{ is odd }, \\
                2(t + k + 1) - |i-j|, & i \neq j \text{ and } |i-j| \text{ is even }.
            \end{cases}\\           
            &\text{ and }\\
            &[\bm{I}_{f}^{(1)}(t, x_1, \ldots, x_{k})]_{ij} =
            \begin{cases}
                0,  & i = j, \\[4pt]
                2t + 1 - |i - j| , & i \neq j \text{ and } |i-j| \text{ is odd },\\
                2t + 2 - |i-j|,    & i \neq j \text{ and } |i-j| \text{ is even }. 
            \end{cases}
            \end{align*}
            
    \end{lemma}
    \begin{proof}
        Consider the set of \( k  \) binary vectors \( \{x_i\}_{i=1}^{k} \) defined by
        \[
            x_i = 
            \begin{cases}
                0^{k-i+1}(10)^{(i-1)/2}, &  \text{ if $i$ is odd},\\[4pt]
                0^{k-i+1}(10)^{(i-2)/2}1, & \text{ if $i$ is even.}
            \end{cases}
        \]
        The function values corresponding to these vectors are given by
        \[
            f_i = f(x_i) = i , \quad \forall\, 1 \leq i \leq k.
        \]
        For any distinct pair \( (x_i, x_j) \), the insertion–deletion distance satisfies
        \[
            d_{ID}(x_i, x_j) =
            \begin{cases}
                |i-j|, & \text{ if $|i-j|$ is even,}\\
                |i - j| + 1, & \text{ if $|i-j|$ is odd.}
            \end{cases}
        \]
        and for any two binary vectors, say $x$ and $y$ of the same length, the insdel distance between the two vectors is bounded below by the modulus of the difference of the number-of-runs of the respective vectors, i.e., $|r(x) -r(y)| \leq d_{ID}(x,y)$. 

        \noindent
        Hence, the corresponding function distance satisfies
        \[
            |i-j| \leq d_{ID}^{f}(f_i, f_j) \leq d_{ID}(x_i, x_j), 
            \quad \forall\, 1 \leq i \neq j \leq k .
        \]
        Since the insdel distance is always even, from the above inequality, we conclude $$d_{ID}^{f}(f_i, f_j) = d_{ID}(x_i, x_j), \quad \forall\, 1 \leq i \neq j \leq k.$$
        Therefore, the entries of the function insdel matrix and insdel distance matrix associated with the set \( \{x_i\}_{i=1}^{k} \) are given respectively by
            
             \begin{align*}
            &[\bm{I}_f^{(2)}(t, f_1, \ldots, f_{k})]_{ij} =
            \begin{cases}
                0,      & i = j, \\[4pt]
                2(t + k) - |i-j| + 1, & i \neq j \text{ and } |i-j| \text{ is odd }, \\
                2(t + k + 1) - |i-j|, & i \neq j \text{ and } |i-j| \text{ is even }.
            \end{cases}\\
            &\text{ and }\\
            &[\bm{I}_{f}^{(1)}(t, x_1, \ldots, x_{k})]_{ij} =
            \begin{cases}
                0,  & i = j, \\[4pt]
                [2t + 1 - |i - j|]^+ , & i \neq j \text{ and } |i-j| \text{ is odd },\\
                [2t + 2 - |i-j|]^+,    & i \neq j \text{ and } |i-j| \text{ is even }. 
            \end{cases}
            \end{align*}      
        This establishes the desired bounds on $r^{f}_{{ID}}(k, t)$, completing the proof.
    \end{proof}
    
    \begin{example}
        Let \( k = 4 \) and \( t = 1 \). Then, from Lemma~\ref{runs_bound}, the function distance matrix corresponding to the number-of-runs function and the insdel distance matrix corresponding to the binary vectors $x_1 = (0000)$, $x_2 = (0001)$, $x_3 = (0010)$ \text{ and } $x_4 = (0101)$ are given respectively by
        \[
            \bm{I}_f^{(2)}(1, f_1, f_2, f_3, f_4) =
            \begin{pmatrix}
                0 & 10 & 10 & 8 \\[4pt]
                10 & 0 & 10 & 10 \\[4pt]
                10 & 10 & 0 & 10 \\[4pt]
                8 & 10 & 10 & 0
            \end{pmatrix},
        \]
        and
        \[
            \bm{I}_{\textit{f}}^{(1)}(1, x_1, x_2, x_3, x_4) =
            \begin{pmatrix}
                0 & 2 & 2 & 0\\[4pt]
                2 & 0 & 2 & 2\\[4pt]
                2 & 2 & 0 & 2\\[4pt]
                2 & 2 & 2 & 0
            \end{pmatrix}.
        \]  
    \end{example}
    The next result presents a lower bound on the redundancy of the number-of-runs function that is inferred from Lemma~\ref{runs_bound} and the unconditional Plotkin-like bound of Lemma~\ref{plotk}.
    
    \begin{corollary}\label{common_lower}
        Let $k \geq t+2$. Then,
        {
            \[
            r_{ID}^f(k,t) \geq \frac{2}{(t+2)^2}\left[\frac{5t(t+1)(t+2)}{6} + \left \lceil\frac{t}{2} \right \rceil (t+1) + \left \lceil\frac{t}{2} \right \rceil^2\right].
            \]
        }
        
        This simplifies to:
        $$ r_{ID}^f(k,t) \geq \frac{10t^3 + 39t^2 + 26t}{3(t+2)^2}, \quad \text{ when $t$ is even, }$$
        and $$ r_{ID}^f(k,t) \geq \frac{10t^3 + 39t^2 + 38t + 29}{3(t+2)^2}, \quad \text{ when $t$ is odd. }$$       
    \end{corollary}
    
    \begin{proof}
         For $1 \leq i \leq k$, let $x_i = 0^{k-i+1}(10)^{(i-1)/2}$ if $i$ is odd and $x_i = 0^{k-i+1}(10)^{(i-2)/2}1$ if $i$ is even. Let $\mathcal{P} = \{p_1, p_2, \ldots, p_k\}$ be an $\bm{I}_f^{(1)}(t, x_1,x_2,\dots,x_k)$-code of length $N^{(1)}_{ID}(\bm{I}_f^{(1)}(t, x_1,x_2,\dots,x_k))$. Consider the first $t+2$ codewords of $\mathcal{P}$ and $(t+2) \times (t+2)$ leading principal submatrix of $\bm{I}_f^{(1)}(t, x_1,x_2,\dots,x_k)$ denoted by $I_{t+2}.$
         Then, $\{p_1, p_2, \ldots, p_{t+2}\}$ is a $I_{t+2}$-code. By Lemma~\ref{plotk} (for $M = t+2$, using $\tfrac{2}{M^2-1} \ge \tfrac{2}{M^2}$ when $M$ is odd, the coefficient $\tfrac{2}{(t+2)^2}$ is valid for both parities of $t+2$),
         {
            \begin{align*}
             N_{ID}^{(1)}(I_{t+2}) &\geq \frac{2}{(t+2)^2}\sum_{i = 1}^{t+2}\sum_{j= i+1}^{t+2}[I_{t+2}]_{ij}\\
                              & =   \frac{2}{(t+2)^2}\left(\sum_{\substack{i=0 \\ i \text{ is even}}}^{t}(t+1 - i)(2t - i) + \sum_{\substack{i=0 \\ i \text{ is odd}}}^{t}(t+1 - i)(2t + 1- i)\right)\\ 
                              & = \frac{2}{(t+2)^2} \left [ \frac{5t(t+1)(t+2)}{6} + \left \lceil\frac{t}{2} \right \rceil (t+1) + \left \lceil\frac{t}{2} \right \rceil^2\right]                             
         \end{align*}
         }
         
         From Lemma~\ref{runs_bound}, we have
         {
            \begin{align*}
             r_{ID}^f(k,t) &\geq N^{(1)}_{ID}(\bm{I}_f^{(1)}(t, x_1,x_2,\dots,x_k))\\[3pt]
             &\geq N_{ID}^{(1)}(I_{t+2})\\[3pt]
             &\geq \frac{2}{(t+2)^2} \left [ \frac{5t(t+1)(t+2)}{6} + \left \lceil\frac{t}{2} \right \rceil (t+1) + \left \lceil\frac{t}{2} \right \rceil^2\right]
         \end{align*}
         }         
         The bounds for the cases of odd and even $t$ are immediate consequences of the preceding bound.
    \end{proof}
    Next, we present a construction of FCIDCs for the number-of-runs function which in turn gives an upper bound on the optimal redundancy of FCIDCs for the number-of-runs function.

    \begin{theorem}
    \label{thm:runs-construction}
    For every $t\ge 1$, the optimal redundancy of FCIDCs for the number-of-runs function $r:\mathbb{F}_2^{k}\to\{1,\dots,k\}$ satisfies
    \[
    r^{r}_{ID}(k,t)\;\le\;N^{(1)}_{ID}\big(2t+1,\,2t+2\big).
    \]
    \end{theorem}

    \begin{proof}
    Let $\{C_0,\dots,C_{2t}\}\subseteq\mathbb{F}_2^{r}$ be a binary insdel code with minimum insdel distance $2t+2$ and length $r=N^{(1)}_{ID}(2t+1,2t+2)$, and define the encoding function
    \[
    \psi(x)=\big(x,\,C_{\,r(x)\bmod(2t+1)}\big).
    \]
    The function-correction requires only those codewords to be $2t+2$ distance apart for which the number of runs value is distinct. We therefore fix $x,y\in\mathbb{F}_2^{k}$ with $r(x)\ne r(y)$ and consider two cases.

    \smallskip
    \noindent\textbf{Case 1: $d_{ID}(x,y)\le 2t$.}
    By Lemma~\ref{run_insdel_bound}, $|r(x)-r(y)|\le d_{ID}(x,y)\le 2t$, and since $r(x)\ne r(y)$ we have $1\le|r(x)-r(y)|\le 2t$. Hence $r(x)\not\equiv r(y)\pmod{2t+1}$, so the two messages receive distinct codewords of $\{C_0,\dots,C_{2t}\}$, at insdel distance at least $2t+2$. By Lemma~\ref{lem:block-distance},
    \[
    d_{ID}(\psi(x),\psi(y))
    \;\ge\; d_{ID}\big(C_{\,r(x)\bmod(2t+1)},\,C_{\,r(y)\bmod(2t+1)}\big)
    \;\ge\; 2t+2.
    \]

    \smallskip
    \noindent\textbf{Case 2: $d_{ID}(x,y)\ge 2t+2$.}
    By Lemma~\ref{lem:block-distance},
    \[
        d_{ID}(\psi(x),\psi(y))\;\ge\;d_{ID}(x,y)\;\ge\;2t+2.
    \]

    \smallskip
    In both cases $d_{ID}(\psi(x),\psi(y))\ge 2t+2$ whenever $r(x)\ne r(y)$, so $\psi$ is a $t$-insdel-error-correcting FCIDC with redundancy $N^{(1)}_{ID}(2t+1,2t+2)$.
    \end{proof}

    \begin{corollary}\label{cor:runs-explicit}
    For every $t\ge 1$ and $k\ge t+2$, the optimal redundancy of FCIDCs for the number-of-runs function $r:\mathbb{F}_2^{k}\to\{1,\dots,k\}$ satisfies
    \[
    r^{r}_{ID}(k,t)\;\le\;
    \max\!\left\{\,2\log_2(2t+1)+12t,\;\;
    \frac{8t}{\ln 2}\,\ln\!\frac{8t}{\ln 2}\,\right\},
    \]
    independently of $k$.
    \end{corollary}

    \begin{proof} By Theorem~\ref{thm:runs-construction}, $r^{r}_{ID}(k,t)\le N^{(1)}_{ID}(2t+1,2t+2)$. Applying Theorem~\ref{N1_M_d_bound} with $M=2t+1$ and $d=2t+2$ (so $d-2=2t$, $4(d-2)=8t$, and $\ln 8=3\ln 2$) yields the explicit bound. 
    \end{proof}

\section{Maximum Run-Length Function}
    Another interesting function that can be defined from the run-length sequence $R(x)$ is the maximum run-length function, which gives the length of the longest block of consecutive identical bits in a given vector.

    \begin{definition}[Maximum Run-Length Function]
        Let $x \in \mathbb{F}_2^k$ and $R(x) = (r_1, r_2, \ldots, r_m)$ then the maximum run-length function is defined as:
        $$r_{\max}(x) = \max_{1 \leq i \leq m}r_i$$
    \end{definition}
    
    \begin{example}
        Consider the vector $u = 0000101$ , then the run-length vector corresponding to the vector $u$ is $(4,1,1,1)$ , and hence the maximum run-length of $u$ is $4.$
    \end{example}

    \noindent The expressiveness for this function is given by $E = |Im(r_{max}(\cdot))| = |\{1,2,\dots,k\}| = k$.

    The following result establishes a lower bound on the function distance for the maximum run-length function.
    \begin{lemma}
    \label{function_dis_lower}
        Let $f: \mathbb{F}_2^k \rightarrow \{1, 2, \ldots,k\}$ be the maximum run-length function. Then,
        \[
            d_{ID}^f(i,j) \geq 2\Big \lceil \frac{|i-j|}{\min(i,j)+1} \Big \rceil \text{ for all } 1 \leq i,j \leq k.
        \]
    \end{lemma}
    \begin{proof}
        Let $x$ be a binary vector of length $k$ and maximum run-length $i$. Without loss of generality, we assume that the longest run of $x$ is $0^i$. To derive the lower bound, we consider a binary vector $y$ of length $k$ and maximum run-length $j$ that is as close as possible to $x$ while respecting the maximum run-length constraint. In this way, the minimal possible insdel distance will occur when the long runs in $x$ and $y$ are of the same symbol. 
        
        Let $y'$ be a substring of $y$ aligned with the run $0^i$. Let \(t\) denote the number of coordinates of \(y'\) equal to \(1\).
    Then \(y'\) contains \(i-t\) zeros. This means $y'$ can have at most $t+1$ blocks of zeros, each split by $1$ and having size at most $j$. Hence, we have,
        \begin{align*}
            i - t \leq (t+1)j \implies t \geq \Big \lceil \frac{i-j}{j+1} \Big \rceil
        \end{align*}
        which means in order to have a run of zeros of length $j$, $y'$ should have at least $\Big \lceil \frac{i-j}{j+1} \Big \rceil$ number of ones.
        Hence, 
        \[
            LCS(x', y') \leq i -\Big \lceil \frac{i-j}{j+1} \Big \rceil
        \]
        as $LCS(x',y')$ is equal to the maximum number of zeros in $y'$. The length of longest common subsequence of $x$ and $y$ can be bounded above in terms of $LCS(x',y')$ as follows:
        \[
            LCS(x,y) \leq LCS(x',y') + (k-i)
        \]
        From the formula of the insdel distance in terms of $LCS$, we get:
        \begin{align*}
            d_{ID}(x,y) &= 2k - 2LCS(x,y)\\
                            &\geq 2k - 2(LCS(x',y') + (k-i)) \\
                            &\geq 2\Big \lceil \frac{i-j}{j+1} \Big \rceil.
        \end{align*}
    \end{proof}

    \begin{remark}
        The bound in Lemma~\ref{function_dis_lower} is tight. For any integers $i, j$ with $0 < j < i \leq k$, there exist binary vectors $x$ and $y$ of length $k$ with $r_{\max}(x) = i$ and $r_{\max}(y) = j$ for which equality is achieved.
    \end{remark}
    \noindent The following example shows that the bound in Lemma~\ref{function_dis_lower} is tight.
    
    \begin{example}
        Let $k = 5$ and $\mathcal{S}= \{00000, 00001, 00010, 00100, 01010\}$. Then for any pair of strings $x,y$ in $\mathcal{S}$ having maximum run-length $i$ and $j$ respectively, $d_{ID}(x, y) = 2\Big \lceil \frac{|i-j|}{\min(i,j)+1} \Big \rceil \text{ for all } 1 \leq i,j \leq 5.$ 
    \end{example}
    
    \noindent Using Lemma~\ref{function_dis_lower} and Corollary~\ref{optional_red_and_function_dis_bound}, we give an upper bound on the optimal redundancy of FCIDCs designed for maximum run-length function.
    
    \begin{lemma}\label{max_runs_upper_bound}
        Let $f(x) = r_{max}(x)$ defined over $\mathbb{F}_2^k$. Then, 
        \[
            r_{ID}^f(k,t) \leq N_{ID}^{(2)}(\bm{I};k).
        \]
        where 
        \[
            [\boldsymbol{I}]_{ij} = 
            \begin{cases}
                0,      & i = j, \\[4pt]
                2\Big (t + k + 1 - \Big \lceil \frac{|i-j|}{\min(i,j) +1} \Big \rceil \Big ) , & i \neq j.                
            \end{cases}
        \]
    \end{lemma} 
    \begin{proof}
        Let $x,y \in \mathbb F_2^k$ be arbitrary vectors such that $f(x)=i$ and $f(y)=j$. By the definition of function distance,
    \[
    d_{ID}(x,y)\ge d^f_{ID}(i,j).
    \]
    Combining this with Lemma~\ref{function_dis_lower}, we obtain
    \[
    d_{ID}(x,y)\ge d^f_{ID}(i,j)
    \ge
    2\left\lceil
    \frac{|i-j|}{\min(i,j)+1}
    \right\rceil.
    \] 
    Let $\boldsymbol{I}$ be a square matrix of order $k$ whose entries are:
        \[
            [\boldsymbol{I}]_{ij} = 
            \begin{cases}
                0,      & i = j, \\[4pt]
                2\Big (t + k + 1 - \Big \lceil \frac{|i-j|}{\min(i,j) +1} \Big \rceil \Big ) , & i \neq j.               
            \end{cases}
        \]
        Then, by Corollary~\ref{optional_red_and_function_dis_bound} $r_{ID}^f(k,t) \leq N_{ID}^{(2)}(\bm{I};k).$       
    \end{proof}

    Observe that $\max_{i,j}[\mathbf{I}]_{ij} = 2(t+k)$ in Lemma~\ref{max_runs_upper_bound}, therefore the following redundancy bound follows from Corollary~\ref{bound_Imax}.

    \begin{corollary}
        Let $k,t \in \mathbb{N}$ and $r_{ID}^f(k,t)$ be the optimal redundancy of FCIDCs corresponding to $f(x) = r_{max}(x)$. Then,
        \[
                r_{ID}^f(k,t) \leq \Bigg \lceil \frac{\ln (k) + 2(t+k-1)\ln \Big(\frac{ek}{t+ k - 1}\Big)}{\ln 2} \Bigg \rceil.
        \]
    \end{corollary}

    The next lemma gives a lower bound on the optimal redundancy of FCIDCs designed for the maximum run-length functions, using Corollary~\ref{lower_bound_red}.
    \begin{lemma}\label{max_lower_bound}
        Let $f(x) = r_{max}(x)$ and $x_i =0^i(10)^{(k-i)/2}$ if $k-i$ is even and $x_i =0^i(10)^{(k-i-1)/2}1$ if $k-i$ is odd for $i \in \{1,2,\ldots,k\}$.Then,
        
          \[
            r_{ID}^f(k,t) \geq N_{ID}^{(1)}(\boldsymbol{I}_f^{(1)}(t, x_1, x_2, \ldots, x_k)).
          \]  
        
        where
        
        \[
            [\boldsymbol{I}_f^{(1)}(t, x_1, x_2, \ldots, x_k)]_{ij}=
            \begin{cases}
                0,  & i = j, \\[4pt]
                2t + 1 - |i - j| , & i \neq j \text{ and } |i-j| \text{ is odd },\\
                2t + 2 - |i-j|,    & i \neq j \text{ and } |i-j| \text{ is even }. 
            \end{cases}
        \]
        
    \end{lemma}
    \begin{proof}
        Consider the following set of representative vectors: $x_i =0^i(10)^{(k-i)/2}$ if $k-i$ is even and $x_i =0^i(10)^{(k-i-1)/2}1$ if $k-i$ is odd for $i \in \{1,2,\ldots,k\}$. Then, $f(x_i) = r_{max}(x_i) = i$. We claim that the insdel distance between these representative set of vectors is given by:
        \[
            d_{ID}(x_i, x_j) =
            \begin{cases}
                |i-j|, & \text{ if $|i-j|$ is even}\\
                |i - j| + 1, & \text{if $|i-j|$ is odd}
            \end{cases}
        \]
        WLOG assume $k$ is even and $i > j$.\\
        \noindent \textbf{Case 1:} Both $i$ and $j$ are even.
        
        \smallskip
        \noindent Then, $LCS(x_i,x_j) \geq j + \frac{i-j}{2} + k- i = k - \frac{i-j}{2}.$\\
        Hence, $d_{ID}(x_i,x_j) = 2(k - LCS(x_i,x_j)) \leq i-j$.\\
        From Lemma~\ref{run_insdel_bound} we have $d_{ID}(x_i,x_j) \geq |r(x_i) - r(x_j)| = |(1 + k -i) -(1 + k - j)| = |j-i| = i-j.$\\
        Hence, $d_{ID}(x_i, x_j) = i-j$.\\
        
        \noindent \textbf{Case 2:} Both $i$ and $j$ are odd.
        
        \smallskip
        \noindent Then, $LCS(x_i,x_j) \geq j + \frac{i-j}{2} + ( k- i-1) +1 = k - \frac{i-j}{2}.$\\
        Therefore, using the same argument as in Case 1, we get $d_{ID}(x_i, x_j) = i-j$.\\
        
        





    \noindent\textbf{Case 3.} $i$ is odd and $j$ is even.\\
    Then, $x_i = 0^i\,(10)^{(\frac{k-i-1}{2})}1$ and $x_j = 0^j\,(10)^{(\frac{k-j}{2})}$.
    Consider a vector $z \;=\; 0^{(i+j-1)/2}\,(10)^{(k-i-1)/2}1 \in \mathbb{F}_2^{k - \frac{i-j+1}{2}}$. We claim that $z$ is a common subsequence of both $x_j$ and $x_i$.
    
    \medskip
    \begin{itemize}
    \item {$z$ is a subsequence of $x_j$:}      
     $z$ can be obtained from $x_j$ by deleting $\frac{i-j+1}{2}$ bits from $x_j$. Delete the first $(i - j - 1)/2$ ones of $x_j$ at positions $j+1, j+3, \ldots, i-2$ and then delete the last zero of $x_j$ at position $k$. After these deletions, $x_j$'s remaining characters at positions $\{1, \ldots, j\} \cup \{j+2, j+4, \ldots, i-1\} \cup \{i, i+1, \ldots, k-1\}$ are
    \[
    \underbrace{0, \ldots, 0}_{j}\;\underbrace{0, \ldots, 0}_{(i-j-1)/2}\;\underbrace{(10)^{(k-i-1)/2}\,1}_{\text{positions } i, \ldots, k-1}
    \;=\; 0^{(i+j-1)/2}\,(10)^{(k-i-1)/2}\,1 \;=\; z.
    \]

    \item {$z$ is a subsequence of $x_i$:}  
    Since $j \leq i - 1$, we have $(i+j-1)/2 \leq i - 1 < i$, so $x_i$'s leading block $0^{i}$ contains strictly more than $(i+j-1)/2$ zeros. Match $z$'s leading $(i+j-1)/2$ zeros with $x_i$'s 0 at positions $1, 2, \ldots, (i+j-1)/2$, and match $z$'s tail $(10)^{(k-i-1)/2}\,1$ with $x_i$'s tail $(10)^{(k-i-1)/2}\,1$ at $x_i$ positions $i+1, i+2, \ldots, k$. The matched positions form the strictly increasing sequence $1, \ldots, (i+j-1)/2, i+1, \ldots, k$ because $(i+j-1)/2 < i + 1$, so $z$ is a subsequence of $x_i$.
    \end{itemize}
    Hence $\mathrm{LCS}(x_i, x_j) \geq k - \frac{i-j+1}{2}$, yielding  
    \[
    d_{ID}(x_i, x_j)  \leq i - j + 1.
    \]  
    By Lemma~\ref{run_insdel_bound}, $d_{ID}(x_i,x_j)\ge |r(x_i)-r(x_j)| = |(1+k-i)-(1+k-j)| = i-j$. Since $(x_i,x_j)$ have equal length, $d_{ID}(x_i,x_j)$ is even, while $i-j$ is odd, we have $d_{ID}(x_i,x_j)\ge i-j+1$. Combined with the upper bound, $d_{ID}(x_i,x_j)=i-j+1$. 

    \smallskip

    \noindent \textbf{Case 4:} $i$ is even and $j$ is odd.\\
    
    Then
    \[
    x_i = 0^{i}\,(10)^{\frac{k-i}{2}}, \qquad 
    x_j = 0^{j}\,(10)^{\frac{k-j-1}{2}}\,1.
    \]

    Consider the word
    \[
    z = 0^{\frac{i+j-1}{2}}\,(01)^{\frac{k-i}{2}} \;\in\; \mathbb{F}_2^{\,k - \frac{i-j+1}{2}}.
    \]
    \begin{itemize}
    \item {$z$ is a subsequence of $x_i$:} 
    Match the leading $\frac{i+j-1}{2}$ zeros of $z$ with the first $\frac{i+j-1}{2}$ zeros of $x_i$ 
    (positions $1,\dots,\frac{i+j-1}{2}$). 
    For the trailing block $(01)^{\frac{k-i}{2}}$, match the $t$-th ``0'' with the zero at position 
    $i+2(t-1)$ of $x_i$ and the following ``1'' with the one at position $i+2t-1$ of $x_i$ 
    ($t=1,\dots,\frac{k-i}{2}$). 
    The matched indices are strictly increasing because $\frac{i+j-1}{2} < i$; hence $z$ is a subsequence of $x_i$.

    \item {$z$ is a subsequence of $x_j$:} 
    Delete from $x_j$ the first $\frac{i-j+1}{2}$ ones (positions $j+1, j+3, \dots, i$). 
    The remaining symbols consist of the zeros of the leading block (positions $1,\dots,j$), 
    the zeros at positions $j+2, j+4, \dots, i-1$ (there are $\frac{i-j-1}{2}$ of them), 
    and the trailing block $(01)^{\frac{k-i}{2}}$ (which starts at position $i+1$ because $i+1$ is odd and is a zero). 
    Concatenating these yields $z$. Thus $z$ is a common subsequence.
    \end{itemize}
    Therefore $\mathrm{LCS}(x_i,x_j) \ge |z| = k - \frac{i-j+1}{2}$, and
    \[
    d_{ID}(x_i,x_j) \le 2\bigl(k - \mathrm{LCS}(x_i,x_j)\bigr) \le i - j + 1.
    \]

    The run counts are $r(x_i)=1+(k-i)$, $r(x_j)=1+(k-j)$, so $|r(x_i)-r(x_j)| = i-j$, which is odd. 
    By Lemma~\ref{run_insdel_bound},
    \[
    d_{ID}(x_i,x_j) \ge i - j + 1.
    \]

    \noindent
    Combining the upper and lower bounds gives $d_{ID}(x_i,x_j) = i - j + 1$.

    \noindent
    Thus, in both Cases 3 and 4, when $i$ and $j$ have opposite parity,  
    \[
        d_{ID}(x_i, x_j) = |i - j| + 1.
    \]
    \[
            [\boldsymbol{I}_f^{(1)}(t, x_1, x_2, \ldots, x_k)]_{ij}=
            \begin{cases}
                0,  & i = j, \\[4pt]
                [2t + 1 - |i - j|]^+ , & i \neq j \text{ and } |i-j| \text{ is odd },\\
                [2t + 2 - |i-j|]^+,    & i \neq j \text{ and } |i-j| \text{ is even }. 
            \end{cases}
        \]
        
        and the lower bound on the redundancy, $r_{ID}^f(k,t) \geq N_{ID}^{(1)}(\boldsymbol{I}_f^{(1)}(t, x_1, x_2, \ldots, x_k))$ follows from Corollary~\ref{lower_bound_red}.
    
    \end{proof}

    \noindent
    Because the matrix entries $\boldsymbol{I}_f^{(1)}(t, x_1, x_2, \ldots, x_k)$(Lemma~\ref{max_lower_bound}) associated with maximum run length function match those for the number-of-runs (Lemma~\ref{runs_bound}), the lower bound on optimal redundancy for FCIDCs is identical in both cases(Corollary~\ref{common_lower}), as given in the corollary below.
    
    \begin{corollary}
        For any $k \geq t+2$. Then,
        {
            \[
            r_{ID}^f(k,t) \geq \frac{2}{(t+2)^2}\left[\frac{5t(t+1)(t+2)}{6} + \left \lceil\frac{t}{2} \right \rceil (t+1) + \left \lceil\frac{t}{2} \right \rceil^2\right].
            \]
        }
        
        This simplifies to:
        $$\displaystyle r_{ID}^f(k,t) \geq \frac{10t^3 + 39t^2 + 26t}{3(t+2)^2}, \quad \text{ when $t$ is even }$$
        and $$\displaystyle r_{ID}^f(k,t) \geq \frac{10t^3 + 39t^2 + 38t + 29}{3(t+2)^2}, \quad \text{ when $t$ is odd. }$$ 
    \end{corollary}

    \begin{example}
    \label{ex:max-runlength-k4-t1}
    Let $k = 4$ and $t = 1$. The representative vectors from Lemma~\ref{max_lower_bound} are
    \[
    x_1 = 0101, \quad x_2 = 0010, \quad x_3 = 0001, \quad x_4 = 0000,
    \]
    with maximum run-lengths $r_{\max}(x_i) = i$ for $i = 1, 2, 3, 4$. The upper-bound matrix $I$ from Lemma~\ref{max_runs_upper_bound} and the lower-bound matrix from Lemma~\ref{max_lower_bound} are, respectively,
    \[
    \bm{I} = 
    \begin{pmatrix}
        0  & 10 & 10 & 8  \\
        10 & 0  & 10 & 10 \\
        10 & 10 & 0  & 10 \\
        8  & 10 & 10 & 0
    \end{pmatrix},
    \qquad
    \bm{I}^{(1)}_{f}(t, x_1, x_2, x_3, x_4) = 
    \begin{pmatrix}
        0 & 2 & 2 & 0 \\
        2 & 0 & 2 & 2 \\
        2 & 2 & 0 & 2 \\
        0 & 2 & 2 & 0
    \end{pmatrix}.
    \]

\end{example}
\section{Locally $(\lambda, \rho)_{ID}$-function}

    Lenz et al.\ first studied a class of functions, called locally binary functions \cite{10132545}, with respect to the Hamming metric and derived optimal function-correcting codes (FCCs) for them. This concept was subsequently generalized by Rajput et al.\ to {locally $(\lambda,\rho)$-functions}\cite{Rajput2025} for the Hamming metric, and later by Verma et al.\ for the $b$-symbol metric \cite{11250740}. The significance of this class of functions stems from the fact that any function with a finite image set can be represented as a locally $(\lambda,\rho)$-function. This universality is the primary reason the theory, initially developed for the Hamming metric, has been extended to other distance metrics where FCCs are being explored. In this section, we study the class of locally $(\lambda,\rho)_{ID}$-functions in the insdel metric setting.

    \begin{definition}[Function ball] The function ball of a function $f: \mathbb{F}_2^k \rightarrow \mathrm{Im}(f)$ with radius $\rho$ around $x \in \mathbb{F}_2^k$ in insdel metric is defined as $$ B_{ID}^f(x,\rho) = f(B_{ID}(x,\rho)) = \{f(y) | y \in B_{ID}(x,\rho)\}.$$    
    \end{definition}

    \begin{definition}[Locally bounded function in insdel metric]
        A locally $(\lambda, \rho)_{ID}$-function $f: \mathbb{F}_2^k \rightarrow \mathrm{Im}(f)$ is a function for which $|B_{ID}^f(x,\rho)| \leq \lambda, \quad \forall x \in \mathbb{F}_2^k. $
    \end{definition}

    The following lemma is a straightforward extension of Lemma~\ref{graph_color_function} to the insdel metric setting. This result has subsequently been employed to derive an upper bound on the optimal redundancy of FCIDCs corresponding to locally $(\lambda, \rho)_{ID}$-functions.

    \begin{lemma}\label{insdel_graph_coloring_function}       
        Let $f : \mathbb{F}_2^k \rightarrow \mathrm{Im}(f)$ be a locally $(\lambda, \rho)_{ID}$-function. 
        Assume that $\mathrm{Im}(f)$ is equipped with a total order $\prec$, and that for every $x \in \mathbb{F}_2^k$, the set $B_{ID}^f(x,\rho)$
        forms a contiguous block with respect to $\prec$.
        Then there exists a mapping
        \[
            \mathrm{Col}_f : \mathbb{F}_2^k \rightarrow [\lambda]
        \]
        such that for all $x,y \in \mathbb{F}_2^k$ satisfying $f(x) \neq f(y) \quad \text{and} \quad d_{ID}(x,y) \leq \rho$, we have $\mathrm{Col}_f(x) \neq \mathrm{Col}_f(y)$.
    \end{lemma}

    \begin{proof}
    Let $E = |\mathrm{Im}(f)|$ and write
    \[
    \mathrm{Im}(f) \;=\; \{y_0 \prec y_1 \prec \cdots \prec y_{E-1}\}
    \]
    for the image of $f$ enumerated in increasing order with respect to $\prec$. Define a coloring $\gamma : \mathrm{Im}(f) \to [\lambda]$ by
    \[
    \gamma(y_j) \;=\; 1 + (j \bmod \lambda).
    \]
    By construction, the colors $\{1, 2, \ldots, \lambda\}$ are assigned cyclically along the order $\prec$. In particular, if $I = \{y_a, y_{a+1}, \ldots, y_{a + m - 1}\}$ is any contiguous block of size $m \leq \lambda$, then $\gamma$ is injective on $I$, since the residues $a, a+1, \ldots, a + m - 1$ modulo $\lambda$ are all distinct.

    Now define $\mathrm{Col}_f : \mathbb{F}_2^k \to [\lambda]$ by
    \[
    \mathrm{Col}_f(x) \;=\; \gamma(f(x)).
    \]
    We claim that $\mathrm{Col}_f$ has the required separation property. Suppose $x, y \in \mathbb{F}_2^k$ satisfy $d_{ID}(x, y) \leq \rho$ and $f(x) \neq f(y)$. Since $d_{ID}(x, x) = 0 \leq \rho$, we have $f(x) \in B^{f}_{ID}(x, \rho)$; and since $d_{ID}(x, y) \leq \rho$, we also have $f(y) \in B^{f}_{ID}(x, \rho)$. Thus, both $f(x)$ and $f(y)$ belong to the same set $B^{f}_{ID}(x, \rho)$, which by hypothesis is a contiguous block of size at most $\lambda$ with respect to $\prec$. Since $\gamma$ is injective on every such block, $\gamma(f(x)) \neq \gamma(f(y))$, hence
    \[
    \mathrm{Col}_f(x) \;=\; \gamma(f(x)) \;\neq\; \gamma(f(y)) \;=\; \mathrm{Col}_f(y),
    \]
    as required.
    \end{proof}

    Throughout the remainder of this section, we assume that all locally $(\lambda, \rho)_{ID}$ -functions under consideration satisfy the assumption in Lemma~\ref{insdel_graph_coloring_function}.
    We now present an upper bound on the optimal redundancy of FCIDCs corresponding to locally $(\lambda, \rho)_{ID}$-functions, expressed in terms of the shortest possible length of a binary insdel code with a prescribed number of codewords and minimum insdel distance.

    \begin{theorem}\label{thm:local_redu_upper}
    Let $t$ be a positive integer and let $f: \mathbb{F}_2^k \rightarrow \mathrm{Im}(f)$ be a locally $(\lambda, 2t)_{ID}$-function. Suppose $\mathrm{Im}(f)$ is equipped with a total order $\prec$ such that for every $x \in \mathbb{F}_2^k$, the function ball
    $B^{f}_{ID}(x, 2t)$ forms a contiguous block with respect to $\prec$. Then
    \[
    r^{f}_{ID}(k, t) \;\leq\; N^{(1)}_{ID}\!\big(\lambda,\, 2t+2\big).
    \]
    \end{theorem}

    \begin{proof}
    By Lemma~\ref{insdel_graph_coloring_function} there is a colouring $\mathrm{Col}_f : \mathbb{F}_2^k \to [\lambda]$ with
    $\mathrm{Col}_f(x) \neq \mathrm{Col}_f(y)$ whenever $f(x)\neq f(y)$ and $d_{ID}(x,y)\leq 2t$. Let $\mathcal{C}=\{C_1,\dots,C_\lambda\}\subseteq \mathbb{F}_2^{r}$ be a binary insdel code of size $\lambda$, minimum insdel
    distance $2t+2$, and length $r = N^{(1)}_{ID}(\lambda, 2t+2)$, and define
    $$\psi(x) = (x,\,C_{\mathrm{Col}_f(x)}).$$
    
    \noindent Let $x,y\in\mathbb{F}_2^k$ with $f(x)\neq f(y)$. 
    
    \noindent If $d_{ID}(x,y)\geq 2t+2$, then Lemma~\ref{lem:block-distance} gives $d_{ID}(\psi(x),\psi(y)) \geq d_{ID}(x,y) \geq 2t+2$.

    \noindent If $d_{ID}(x,y)\leq 2t$, then $\mathrm{Col}_f(x)\neq\mathrm{Col}_f(y)$, so $C_{\mathrm{Col}_f(x)}$ and $C_{\mathrm{Col}_f(y)}$ are distinct codewords of $\mathcal{C}$ with $d_{ID}\big(C_{\mathrm{Col}_f(x)},C_{\mathrm{Col}_f(y)}\big)\geq 2t+2$,
    and Lemma~\ref{lem:block-distance} gives 
    \[
    d_{ID}(\psi(x),\psi(y)) \geq d_{ID}\big(C_{\mathrm{Col}_f(x)},C_{\mathrm{Col}_f(y)}\big)\geq 2t+2.
    \]
    \noindent In both cases $\psi$ is an FCIDC with redundancy $N^{(1)}_{ID}(\lambda, 2t+2)$.
    \end{proof}
    
    \begin{remark}
    \label{rmk:scope-of-thm74}
    The requirement that $\mathrm{Im}(f)$ admits a total order and every function ball forms a contiguous block is a restriction, but it is satisfied by several natural functions. For example, the image set $\{1,\dots,k\}$ of the number-of-runs function of Section~\ref{sec:runs} admits the natural order. Lemma~\ref{run_insdel_bound} shows that every two distinct function values of the number-of-runs function that are within $2t$ insdel distance differ by a value of at most $2t$. Thus, the function ball forms an interval and, hence, is contiguous. This allows us to apply Theorem~\ref{thm:local_redu_upper} directly to the number-of-runs function.
    \end{remark}

    We now examine the two smallest instances of $\lambda$, for which the redundancy code can be given explicitly. The instance $\lambda=2$ represents the insdel equivalent of the locally binary functions discussed in \cite{10132545}.
    
    \begin{corollary}
    Let $t$ be a positive integer and $f: \mathbb{F}_2^k \rightarrow \mathrm{Im}(f)$ a locally $(2, 2t)_{ID}$-function with $|\mathrm{Im}(f)| \geq 2$. Then
    \[
    t \;\leq\; r_{ID}^f(k,t) \;\leq\; t+1.
    \]
    \end{corollary}
    
    \begin{proof}
    The lower bound follows from Corollary~\ref{lower_bound_red}. For the upper bound, $0^{t+1}$ and $1^{t+1}$ have insdel distance $2(t+1)=2t+2$, so $N^{(1)}_{ID}(2,2t+2)=t+1$, and Theorem~\ref{thm:local_redu_upper} applies.
    \end{proof}

    \begin{corollary}
    \label{local-3-upper}
    Let $t$ be a positive integer and $f: \mathbb{F}_2^k \rightarrow \mathrm{Im}(f)$ a locally $(3, 2t)_{ID}$-function whose image admits a total order making every ball $B^f_{ID}(u,2t)$ contiguous. Then $r^{f}_{ID}(k, t) \leq 2t+2$.
    \end{corollary}
    \begin{proof}
    The codewords $(00)^{t+1},(01)^{t+1},(11)^{t+1}$ of length $2(t+1)$ satisfy $LCS((00)^{t+1},(01)^{t+1})=LCS((11)^{t+1},(01)^{t+1})=t+1$ and $LCS((00)^{t+1},(11)^{t+1})=0$, so their pairwise insdel distance is at least $2t+2$. Hence, $N^{(1)}_{ID}(3,2t+2)\leq 2t+2$ and Theorem~\ref{thm:local_redu_upper} gives the upper bound.
    \end{proof}

    The next theorem applies the Plotkin-like bound on three pairwise-close messages, all mapping to distinct function values, to establish a lower bound on the optimal redundancy of the locally $(3, 2t)_{ID}$-function.
    
    \begin{theorem}
    \label{thm:local-3-lower}
    Let $t \in \mathbb{N}$ and let $f : \mathbb{F}_2^k \to \mathrm{Im}(f)$ be a locally $(3, 2t)_{ID}$-function with $|\mathrm{Im}(f)| \geq 3$. Suppose there exist $x_1, x_2, x_3 \in \mathbb{F}_2^k$ with $f(x_i) \neq f(x_j)$ for all $i \neq j$ and pairwise insdel distances $d_{ID}(x_i, x_j) = 2$. Then
    \[
    r^{f}_{ID}(k, t) \;\geq\; \left\lceil \frac{3t}{2} \right\rceil.
    \]
    \end{theorem}

    \begin{proof}
    By Corollary~\ref{lower_bound_red}, $r^{f}_{ID}(k, t) \geq N^{(1)}_{ID}\!\big(I^{(1)}_{f}(t, x_1, x_2, x_3)\big)$. 
    The type~1 insdel-distance matrix has entries
    \[
    \big[I^{(1)}_{f}(t, x_1, x_2, x_3)\big]_{ij} 
    \;=\; \big[\,2t + 2 - d_{ID}(x_i, x_j)\,\big]^{+} 
    \;=\; 2t \qquad \text{for } i \neq j,
    \]
    so
    \[
    I^{(1)}_{f}(t, x_1, x_2, x_3) 
    \;=\; \begin{pmatrix} 0 & 2t & 2t \\ 2t & 0 & 2t \\ 2t & 2t & 0 \end{pmatrix}.
    \]
    By the Plotkin-like bound for irregular insdel-distance codes (Lemma~\ref{plotk}), with $M = 3$ odd,
    \[
    N^{(1)}_{ID}\!\big(I^{(1)}_{f}(t, x_1, x_2, x_3)\big) 
    \;\geq\; \frac{2}{M^2 - 1} \sum_{i < j} \big[I^{(1)}_{f}\big]_{ij} 
    \;=\; \frac{2}{8} \cdot 6t 
    \;=\; \frac{3t}{2}.
    \]
    Since $N^{(1)}_{ID}\!\big(I^{(1)}_{f}(t, x_1, x_2, x_3)\big)$ is a non-negative integer, $N^{(1)}_{ID}\!\big(I^{(1)}_{f}(t, x_1, x_2, x_3)\big)  \geq \lceil 3t / 2 \rceil$, and therefore
    \[
    r^{f}_{ID}(k, t) \;\geq\; \left\lceil \frac{3t}{2} \right\rceil. 
    \]
    \end{proof}

    Together with the corollary~\ref{local-3-upper}, the above theorem gives the bound $\lceil 3t/2\rceil \le r^{f}_{ID}(k,t)\le 2t+2$ on the optimal redundancy of any locally $(3,2t)_{ID}$-function that admits three distinct messages, each corresponding to different function values and possessing a pairwise insdel distance of $2$.
    
    Given that any function having a finite image set can be expressed as a locally $(\lambda, \rho)_{ID}$-function, we next analyze the number-of-runs functions 
    as locally $(\lambda, 2t )_{ID}$-bounded functions.
    
    \begin{proposition}
    \label{prop:runs-local}
        Let $t \in \mathbb{N}$ then the number-of-runs function is a locally $(4t+1, 2t)_{ID}$-function, i.e., $$|B^f_{ID}(x,2t)| \leq 4t + 1, \quad \forall x \in \mathbb{F}_2^k$$
    \end{proposition}
    \begin{proof}
        Let $f = r(x)$ be a number-of-runs function.  
        We claim that
        \[
             |B^f_{ID}(x,2t)| \leq 4t + 1.
        \]
        
        Suppose, for the sake of contradiction, that there exists some $x \in \mathbb{F}_2^k$ such that  
        \[
            |B_{ID}^f(x,2t)| > 4t + 1.
        \]
        Then we can find two vectors $y_1, y_2 \in \mathbb{F}_2^k$ with  
        $d_{ID}(x,y_1) \le 2t$, $d_{ID}(x,y_2) \le 2t$, and 
        \[
            f(y_1) = \max B_{ID}^f(x,2t), \qquad 
            f(y_2) = \min B_{ID}^f(x,2t).
        \]
        Clearly,
        \begin{equation}\label{f_l}
            f(y_1) - f(y_2) \ge  4t + 1.
        \end{equation}
        On the other hand, using the results of Lemma~\ref{run_insdel_bound} and the triangle inequality, we get,
        \begin{equation}\label{f_u}
            {r}(y_1) - r(y_2) 
           \le d_{ID}(y_1,y_2) 
           \le d_{ID}(x,y_1) + d_{ID}(x,y_2) 
           \le 4t.
        \end{equation}
        Thus from equation~\eqref{f_l} and \eqref{f_u} we have,
        \[
            4t \ge f(y_1) - f(y_2) \ge  4t + 1.
        \]
        which is a contradiction.  \\        
        Therefore, 
        \[
            |B_{ID}^f(x,2t)| \le 4t + 1
            \qquad \text{for all } x \in \mathbb{F}_2^k.
        \]
        Hence, the function $f$ is locally $(4t+1, 2t)_{ID}$-function.
    \end{proof}

   \begin{remark}
    \label{rmk:runs-local-vs-tailored}
    Proposition~\ref{prop:runs-local} shows that the number-of-runs function is locally $(4t+1, 2t)_{ID}$, so Theorem~\ref{thm:local_redu_upper} applies and yields the upper bound
    \[
    r^{r}_{ID}(k,t) \;\le\; N^{(1)}_{ID}(4t+1,\, 2t+2).
    \]
    This is, however, weaker than the bound
    \[
    r^{r}_{ID}(k,t) \;\le\; N^{(1)}_{ID}(2t+1,\, 2t+2)
    \]
    obtained for the same function in Theorem~\ref{thm:runs-construction}, since $N^{(1)}_{ID}(M,d)$ is non-decreasing in $M$ and $2t+1 < 4t+1$. Thus, although the locally-bounded framework applies, the tailored construction gives the sharper estimate for the number-of-runs function.
    \end{remark}

\begin{table}[ht]
    \centering
    \begin{threeparttable}
    \renewcommand{\arraystretch}{2.2}
    \setlength{\tabcolsep}{5pt}
    \begin{tabular}{|>{\raggedright\arraybackslash}p{0.19\textwidth}|
                    >{\centering\arraybackslash}p{0.32\textwidth}|
                    >{\centering\arraybackslash}p{0.39\textwidth}|}
        \hline
        \textbf{Function} & \textbf{Lower Bound} & \textbf{Upper Bound} \\
        \hline
        \multirow{2}{=}{VT-syndrome function}
            &{$\dfrac{2tk}{k+1}$}
            & \multirow{2}{=}{$\left\lceil \dfrac{\ln (k+1) + 2(t+k-1)\ln\!\left(\frac{ek}{t+k-1}\right)}{\ln 2} \right\rceil$} \\
            & & \\
        \hline
        \multirow{2}{=}{number-of-runs function}
            & $\dfrac{10t^3 + 39t^2 + 26t}{3(t+2)^2}$ \quad ($t$ even)
            & \multirow{2}{=}{$\max\!\left\{2\log_2(2t+1)+12t,\ \dfrac{8t}{\ln 2}\ln\dfrac{8t}{\ln 2}\right\}$} \\
        \cline{2-2}
            & $\dfrac{10t^3 + 39t^2 + 38t + 29}{3(t+2)^2}$  ($t$ odd)
            & \\
        \hline
        \multirow{2}{=}{Maximum run-length function}
            & $\dfrac{10t^3 + 39t^2 + 26t}{3(t+2)^2}$ \quad ($t$ even)
            & \multirow{2}{=}{$\left\lceil \dfrac{\ln (k) + 2(t+k-1)\ln\!\left(\frac{ek}{t+k-1}\right)}{\ln 2} \right\rceil$} \\
        \cline{2-2}
            & $\dfrac{10t^3 + 39t^2 + 38t + 29}{3(t+2)^2}$  ($t$ odd)
            & \\
        \hline
        Locally $(2,2t)_{ID}$-function
            & $t$
            & $t+1$ \\
        \hline
        Locally $(3,2t)_{ID}$-function
            & $\left\lceil \dfrac{3t}{2} \right\rceil$\tnote{*} 
            & $2t+2$ \\
        \hline
    \end{tabular}
    \begin{tablenotes}
    \footnotesize
    \item[*] Holds for locally $(3,2t)_{ID}$-functions admitting three messages with pairwise distinct values and pairwise insdel distance $2$; see Theorem~\ref{thm:local-3-lower}.
    \end{tablenotes}
    \end{threeparttable}
    \caption{Lower and upper bounds on the optimal redundancy of FCIDCs for different functions.}
    \label{tab:redundancy-summary}
\end{table}

\section{Conclusion}
This work introduces a comprehensive, unified framework for function-correcting codes specifically designed for insertion–deletion (insdel) channels. A central finding is that, unlike the Hamming, symbol-pair, and Lee metrics, the insdel metric is not additive across a systematic split. The cross-block alignment between concatenated words can reduce the insdel distance of a concatenation by up to $2\min\{k,r\}$. To handle this loss, we introduce two types of insdel-distance matrices and show that nontrivial systematic function-correcting insdel codes are confined to the regime $r \ge k$, since for $r < k$ the message vectors must already form a classical $t$-insdel code and the framework offers no savings. To characterize the fundamental limits of these codes, we define irregular insdel-distance codes, relate their lengths to the optimal redundancy $r^{f}_{ID}(k,t)$, and establish Gilbert–Varshamov-type and Plotkin-like bounds for them. As insdel balls are centre-dependent, these bounds rest on supersequence and run-based arguments rather than uniform sphere-counting. By exploiting the relationship between optimal redundancy and the minimal lengths of irregular insdel-distance codes, we derive a simplified, broadly applicable lower bound on redundancy. Building on these theoretical insights, we derive bounds and explicit codes for several insdel-native function families, including the VT-syndrome function, the number-of-runs function, the maximum-run-length function, and locally $(\lambda, \rho)_{ID}$-bounded functions. For the number-of-runs function, we give a dedicated residue-class construction that improves on the generic locally-bounded estimate, and for locally $(2,2t)_{ID}$- and $(3,2t)_{ID}$-functions, we obtain matching-order lower and upper bounds.

Several directions remain open. Extending function-error correction to DNA data storage, where insertions and deletions are particularly prevalent, and run-related quantities govern synthesis and sequencing reliability, is a natural and promising application. On the theoretical side, tightening the closed-form upper bounds for insdel-native functions, characterizing the lower bound for general locally $(\lambda, 2t)_{ID}$-functions, and extending the framework to non-binary alphabets are appealing avenues for future work.


\section*{Acknowledgment}
We are grateful to Prof. Eitan Yaakobi for his valuable discussions and suggestions, which helped improve the current version of the manuscript.

\printbibliography
\end{document}